\documentclass[twocolumn,nofootinbib,floatfix,showpacs,amsmath,amssymb,pra]{revtex4}
\usepackage{times}
\usepackage{graphicx}
\usepackage{pstricks}
\usepackage{dcolumn}
\usepackage{bm}
\usepackage{bbm} 

\def\s0#1#2{\mbox{\small{$ \frac{#1}{#2} $}}}
\def\0#1#2{\frac{#1}{#2}}
\newcommand{\eq}[1]{(\ref{eq:#1})}
\newcommand{\Eq}[1]{Eq.~(\ref{eq:#1})}

\newcommand{\Fig}[1]{Fig.~\ref{fig:#1}}
\newcommand{\fig}[1]{\ref{fig:#1}}

\newcommand{\Sect}[1]{Sect.~\ref{sec:#1}}

\newcommand{\xv}{\vec{x}}
\newcommand{\Tr}{\mbox{Tr$\,$}}
\newcommand{\cN}{\mathcal{N}}

\newcommand{\ead}[3]{\ensuremath{\Gamma_{#2,#3}^{(#1)}}}
\newcommand{\pdt}{\ensuremath \partial_{\tau}}
\newcommand{\grg}[2][\tau]{\ensuremath{(G_{#1}[\partial_{#1} R_{#1}]G_{#1})_{#2}}}
\newcommand{\pr}[2]{\ensuremath{G_{#1,#2}}}
\newcommand{\texteadi}[2]{\ensuremath{[ \Gamma_{#1}^{(2)} ]^{-1}_{#2}}}
\newcommand{\eadi}[2]{\ensuremath{\left[ \Gamma_{#1}^{(2)} \right]^{-1}_{#2}}}
\newcommand{\eaditext}[2]{\ensuremath{[ \Gamma_{#1}^{(2)} ]^{-1}_{#2}}}
\newcommand{\ea}[1][\tau]{\ensuremath{\Gamma_{#1}}}
\newcommand{\cf}[2][x]{\ensuremath{\phi_{#2}(#1_{#2})}}
\newcommand{\cii}[2][t_0]{\ensuremath{\int_{\mathcal{C},#2}}}
\newcommand{\eado}[2]{\ensuremath{\Gamma_{#2}^{(#1)}}}
\newcommand{\ci}[3][t_0]{\ensuremath{\int_{\mathcal{C}(#2),#3}}}
\newcommand{\cdf}[2]{\ensuremath{\delta_{\mathcal{C}}(x_{#1}-x_{#2})}}
\newcommand{\prw}[4]{\ensuremath{G_{#1,#2}(#3,#4)}}
\newcommand{\tead}[3]{\ensuremath{\Gamma_{#2,#3}^{(#1)s}}}
\newcommand{\tv}[3]{\ensuremath{\Gamma_{#1}^{(4)s}(x_{#2},x_{#3})}}
\newcommand{\cfi}[3]{\ensuremath{I_{#1}(x_{#2},x_{#3})}}
\newcommand{\priw}[4]{\ensuremath{G^{-1}_{#1,#2}(#3,#4)}}
\newcommand{\cdfw}[2]{\ensuremath{\delta_{\mathcal{C}}^{#1 #2}(x_{#1}-x_{#2})}}
\newcommand{\tht}[1]{\ensuremath{\theta(t-t_{#1})}}
\newcommand{\thf}[2]{\ensuremath{\theta(t_{#1}-t_{#2})}}
\newcommand{\ff}[4]{\ensuremath{F_{#1,#2}(x_{#3},x_{#4})}}
\newcommand{\rf}[4]{\ensuremath{\rho_{#1,#2}(x_{#3},x_{#4})}}
\newcommand{\sgn}[2]{\ensuremath{\mbox{sgn}_{\mathcal{C}}(t_{#1} - t_{#2})}}
\newcommand{\cfif}[3]{\ensuremath{I_{#1}^F(x_{#2},x_{#3})}}
\newcommand{\cfir}[3]{\ensuremath{I_{#1}^{\rho}(x_{#2},x_{#3})}}
\newcommand{\cfid}[4]{\ensuremath{I^{#1}_{#2}(x_{#3},x_{#4})}}
\newcommand{\spfr}[3]{\ensuremath{ F_{#1}^{2}(x_{#2},x_{#3}) - \frac{1}{4} \rho_{#1}^{2}(x_{#2},x_{#3}) }}

\begin{document}

\title{Far-from-equilibrium quantum many-body dynamics}
\author{Thomas Gasenzer}
\thanks{email:T.Gasenzer@uni-heidelberg.de}
\author{Stefan Ke\ss ler}
\thanks{Present address: Department of Physics, Center for NanoScience, and Arnold Sommerfeld Center 
for Theoretical Physics, Ludwig-Maximilians-Universit\"at M\"unchen, Theresienstr.~37, D-80333 M\"unchen, Germany} 
\author{Jan M. Pawlowski}
\thanks{email:J.Pawlowski@thphys.uni-heidelberg.de}
\affiliation{Institut f\"ur Theoretische Physik, Universit\"at Heidelberg,~Philosophenweg 16, 69120 Heidelberg, Germany}
\affiliation{ExtreMe Matter Institute EMMI,
             GSI Helmholtzzentrum f\"ur Schwerionenforschung GmbH, 
             Planckstra\ss e~1, 
             64291~Darmstadt, Germany} 

\begin{abstract}
\noindent 
The theory of real-time quantum many-body dynamics as put forward in Ref.~\cite{Gasenzer:2008zz} is evaluated in detail. 
It is based on a generating functional of  correlation functions where the closed time contour extends only to a given time. 
Expanding the contour from this time to a later time leads to a dynamic flow of the generating functional. 
This flow describes the dynamics of the system and has an explicit causal structure. 
In the present work it is evaluated within a vertex expansion of the effective action leading to time evolution equations for Green functions.  
These equations are applicable for strongly interacting systems as well as for studying the late-time behaviour of nonequilibrium time evolution. 
For the specific case of a bosonic $\mathcal{N}$-component $\phi^4$-theory with contact interactions an $s$-channel truncation is identified to yield equations identical to those derived from the 2PI effective action in next-to-leading order of a $1/\mathcal{N}$ expansion. 
The presented approach allows to directly obtain non-perturbative dynamic equations beyond the widely used 2PI approximations.
\end{abstract}
\pacs{03.75.Kk, 03.75.Nt, 05.30.-d, 05.70.Ln, 11.15.Pg
\hfill HD--THEP--10--02}

\maketitle
\section{Introduction}
\label{sec:intro}
In recent years quantum field theoretical progress has been achieved towards a reliable description of the dynamics far from equilibrium.
At the same time, experimental efforts to measure the dynamical evolution of complex quantum systems in a time-resolved way have multiplied.
In particular ultracold atomic quantum gases provide a new and unique way to induce and observe many-body time evolution far from thermal equilibrium.
Of primary interest and a fundamental character thereby are experiments setups which allow to induce strong correlations in the system.
These include, e.g., the time evolution of Bose- and Fermi gases in low-dimensional optical lattice traps \cite{Fertig2005a}, studies of equilibration in one-dimensional trapping geometries and the role of integrability of the system under consideration \cite{Kinoshita2006a,Hofferberth2007a,Ritter2007a}.
Recently advanced experimental system preparation and imaging techniques promise substantial progress in the study of time-resolved many-body dynamics \cite{Hofferberth2007a,Esteve2008a,Gericke2008a,Bakr2009a,Karski2009a}.
Many interesting problems in non-equilibrium dynamics have recently been studied in other areas like solid-state physics, heavy-ion collisions, or cosmology.
These include, e.g., strongly correlated driven situations as transport through quantum dots \cite{Heinzel2006a}, the dynamics of (quantum) phase transitions \cite{Hohenberg1977a}, instabilities in a quark-gluon plasma \cite{Strickland:2007fm}, or turbulence in reheating after inflation \cite{Micha:2004bv}.

\begin{figure}[tb]
\begin{center}
\vspace*{-1ex} 
\includegraphics[width=0.43\textwidth]{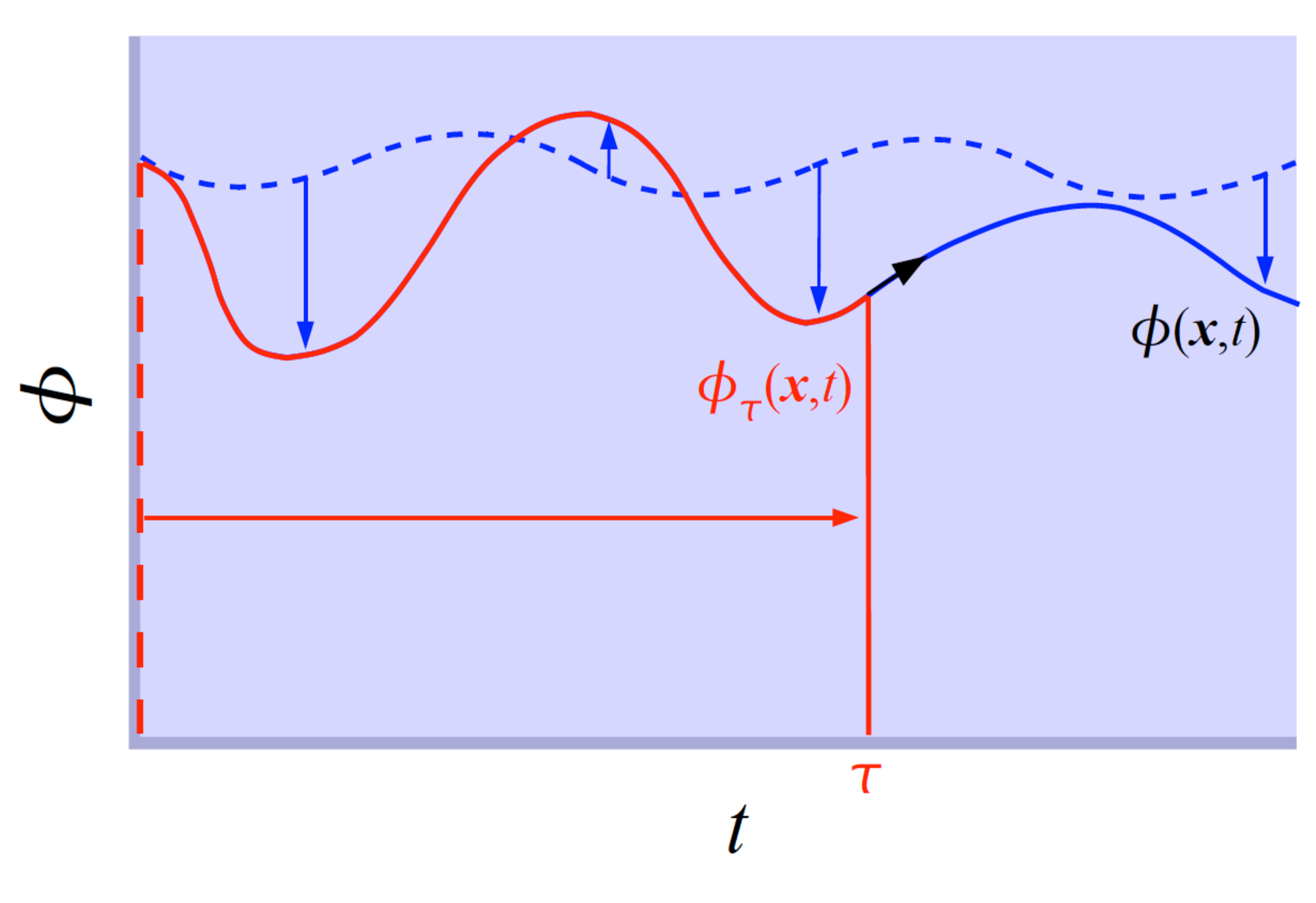}
\end{center}
\vspace*{-3ex} 
\caption{(Color online) 
The time evolution of an interacting quantum field $\phi(\mathbf{x},t)$ (solid line), sketched at some point $\mathbf{x}$, can be determined in different ways: 
By direct integration of an integro-differential equation of motion (black arrow), or by the evolution of flow equations which, e.g., turn the evolution of a non-interacting field (blue dashed line) into that of the interacting one (blue arrows).
In this article we consider explicitly the causal flow of solutions of the interacting field equations up to some time $\tau$ under a change of $\tau$ (red arrow).
The flow-equation approach described here can be used to determine $\phi(\mathbf{x},t)$ in any of the above ways.}
\label{fig:Sketch}
\end{figure}
In this paper we derive dynamic equations for quantum fields far from equilibrium, extending our work presented in Ref.~\cite{Gasenzer:2008zz}. 
We obtain a new time evolution equation for the non-equilibrium effective action by use of functional renormalization group (RG) techniques, cf.~Refs.~\cite{Wetterich:1992yh,Bagnuls:2000ae,Litim:1998nf,Pawlowski:2005xe}, as well as \cite{Canet:2003yu} for non-equilibrium applications.
This exact and closed evolution equation allows for non-perturbative approximations that lead substantially beyond mean-field and quantum Boltzmann type approaches: it can be rewritten, in a closed form, as a hierarchy of dynamical equations for Green functions, and this hierarchy admits truncations that neither explicitly nor implicitly rely on small bare couplings or close-to-equilibrium evolutions.  
The hierarchy of equations is technically close to 2PI and Dyson-Schwinger type evolutions, see \cite{Pawlowski:2005xe}, as well as to evolution equations for the effective action as derived in \cite{Wetterich:1996ap}.
The RG-inspired methods presented here need to be distinguished from the time-dependent DMRG techniques building on quantum information theory which have been intensely studied in the recent past \cite{Vidal2004a}.

As compared to descriptions in frequency space, which are commonly employed to obtain, e.g., non-equilibrium information from spectral densities, real-time approaches profit from causality.
Causality simplifies the practical solution of complex integro-differential equations of motion as integrals in these equations never involve integrands evaluated in the future, where they are unknown.
In the derivation of a dynamical many-body theory on the basis of functional renormalization group techniques presented in this paper we profit from causality in a two-fold way:
We first derive the structure of the equations which depends on the degree to which interaction-induced correlations are taken into account, by means of a flow-equation approach.
By choosing a cut-off function which suppresses fluctuations of the quantum fields at times later than the flow parameter $\tau$ one makes use of causality to derive directly non-perturbative integro-differential equations which effectively resum higher-order correlations as relevant up to the maximum time set by $\tau$.
Causality then plays a second decisive role in giving a Volterra-like structure to the resulting equations, allowing them to be solved in an iterative way without self-consistency complications between the integrands and the solution of the differential propagation. 

Our use of the flow equations to derive dynamic equations valid up to a given cut-off time represents one possibility.
As sketched in \Fig{Sketch} it is alternatively possible to derive the dynamical process through a flow from a solution for a non-interacting system to the full interacting evolution.

Functional quantum-field-theoretical techniques which are related to the approach presented here and profit from causality have been developed and applied intensely in recent years \cite{Berges:2001fi,Aarts:2002dj,Berges:2004yj,Mihaila:2003mh,Berges:2002wr,Gasenzer:2005ze,Aarts:2006cv,Berges2007a,Gasenzer2009a}.
These are based on the two-particle irreducible (2PI) effective action \cite{Luttinger1960a,Baym1962a,Cornwall1974a} which is a functional of the time-ordered two-point correlation or Green functions like $\langle\mathcal{T}\Psi^\dagger(\mathbf{x},t)\Psi(\mathbf{y},t')\rangle$. 
Equations of motion for these correlation functions are derived from the action by means of Hamilton's variational principle.
This procedure requires an explicit expression for the action functional which, in turn, is given by the series of all 2PI `vacuum', i.e., external-leg-less diagrams which can be written down for a given microscopic model.

Strong correlations in general require non-perturbative approximations. 
For the 2PI effective action of an $O(\mathcal{N})$-symmetric theory an expansion in the inverse of the number of field components $\mathcal{N}$ has been studied which in practice is obtained by resumming an infinite number of 2PI diagrams belonging to a particular class \cite{Berges:2001fi,Aarts:2002dj,Berges:2004yj,Gasenzer2009a,Scheppach:2009wu}. 
In this article we show explicitly that one recovers the dynamic equations for such a model, in next-to-leading order of the $1/\mathcal{N}$ expansion, without the need of a resummation procedure, from our functional flow-equation approach \cite{Gasenzer:2008zz}. 
Beyond this result our approach provides a means to derive non-perturbative approximations beyond the large-$\mathcal{N}$ expansion and to check the self-consistency of these truncations without the need to compute the next order of approximation.
Large-$\mathcal{N}$ approximations play an increasing role in experiments with ultracold atoms.
Strong interest is presently arising in the context of the alkaline-earth elements in which the nuclear and electronic spins can be almost perfectly decoupled.
For this reason, the scattering lengths characterizing the interactions in and between the $\mathcal{N}$ magnetic hyperfine levels are essentially identical, implying a model Hamiltonian with $SU(\mathcal{N})$ symmetry \cite{Gorshkov2009a}.
One has $\mathcal{N}=2I+1$, which can be  as large as 10, providing a well-controlled setting for description by as well as testing of $1/\mathcal{N}$ expansions.
The Anderson impurity and Kondo-lattice models studied in this context have been developed in the context of heavy-fermion metals and are thus relevant for a large set of applications.

Our paper is organized as follows:
In \Sect{FRG} we motivate our approach and recall basics of the functional-integral description of time-dependent quantum field theory.
\Sect{RGDynamics} gives the implementation of our approach in terms of functional integrals over a variable closed time (Schwinger-Keldysh) contour and details of the derivation of the dynamical flow equation for the effective action.
In \Sect{FlowEqs} we derive the flow equations for the proper $n$-point functions, $n\leq4$, which form the $n$th moments of the action functional and provide the grounds of the derivation of a coupled set of dynamic integro-differential equations for the field expectation value and the connected two-point correlation function or propagator in \Sect{Dynamics}.
In \Sect{NonPertDyn} we describe in detail an $s$-channel truncation of the resulting dynamic equations and their equivalence to the equations derived from the 2PI effective action in next-to-leading order of a large-$\mathcal{N}$ expansion.
\Sect{Concl} contains our conclusions, and appendices provide technical details of the derivations.

\section{Non-equilibrium quantum field theory}
\label{sec:FRG}
%
\subsection{Motivation}
\label{sec:PTvsNPT}
We consider a system of interacting identical bosons which can be described
by the Hamiltonian  
\begin{align}
\label{eq:MBHamiltonian}
 &H
 =\int \mathrm{d}^dx\Big[\Phi^\dagger(\xv)H_\mathrm{1B}(\xv)\Phi(\xv)
 \nonumber\\
 &\qquad
 +\ \frac{g}{2}
     \Phi^\dagger(\xv)\Phi^\dagger(\xv)
     \Phi(\xv)\Phi(\xv)
    \Big],
\end{align}
with the one-body Hamiltonian
\begin{align}
 &H_\mathrm{1B}(\xv)
 = -\frac{\hbar^2}{2m}\nabla^2_{\xv} + V_\mathrm{trap}(\xv),
\label{eq:1BHamiltonian}
\end{align}
where the field operators obey the usual bosonic commutation relations, $V_\mathrm{trap}(\xv)$ denotes a possible trapping potential, and $g$ the coupling constant which, e.g., in $d=3$ dimensions is given in terms of the $s$-wave scattering length $a$ as $g=4\pi\hbar^2a/m$.
Our approach presented in the following can be applied straightforwardly to the case of fermions.

The dynamics of this many-body system can be described in the Schr\"odinger picture, where the many-body state at time $t$ is given  by some density matrix $\rho_D(t)$. 
The Schr\"odinger or Liouville equation $i\partial_{t}\rho_{D}(t)=[H,\rho_{D}(t)]/\hbar$, with the Hamiltonian (\ref{eq:MBHamiltonian}), determines the dynamical evolution of this state. 
All information about the quantum theory can be encoded in the infinite set of correlation or $n$-point functions
\begin{align}
\label{eq:CorrFunc}
  \langle \Phi^\dagger(x_1)\cdots\Phi(x_n)\rangle_t
  &= \Tr[\rho_D(t)\Phi^\dagger(x_1)\cdots\Phi(x_n)],
\end{align}
with $n \ge 1$. For the Hamiltonian (\ref{eq:MBHamiltonian}), the equation of motion of any particular $n$-point function involves other correlation functions up to the order of $n+2$.
Hence, the result is an infinite system of coupled differential equations for the correlation functions, in analogy to the BBGKY hierarchy in statistical physics. 
This system can in general not be solved exactly and requires the choice of a suitable approximation scheme. 

A commonly used leading-order approximation scheme is based on rewriting the BBGKY hierarchy in terms of cumulant or connected correlation functions and neglecting all cumulants of order three and higher.
The resulting set of equations for the connected (one- and) two-point functions constitutes the mean-field approximation which is generically a reliable approximation for short-time evolutions and/or small couplings:
Larger couplings imply shorter evolution times over which mean-field approximations are valid.
Beyond the mean-field level the BBGKY approach yields, in leading order, to quantum-Boltzmann type approximations.

In deriving dynamic equations beyond the mean-field approximation care is needed in hindsight of conservation laws.
Crucial quantities like the energy-momentum tensor or the particle density-current need to be conserved within the approximation of choice as otherwise secularity problems may arise and the applicability range of the resulting equations be strongly reduced.
Functional-integral techniques as applied in the following provide means to ensure such conservation laws within non-perturbative approximations far beyond mean-field and quantum-Boltzmann approximations.

\subsection{Functional-integral approach}
\label{sec:NeqQFT}
%
\subsubsection{Generating functional}
\label{sec:GenFunc}
Before introducing our functional renormalization-group approach to nonequilibrium dynamics we would like to recall, for the purpose of making our article sufficiently self-contained, some basics about nonequilibrium quantum field theory.
For more details, cf., e.g., Refs. \cite{Berges:2004yj, Rammer2007a,Gasenzer2009a}.
All information about a non-equilibrium quantum many-body system is contained in the generating functional for non-equilibrium correlation functions %
\footnote{Here and in the following we use natural units where $\hbar=1$.}:
\begin{align}
\label{eq:NEgenFuncZ}
  Z[J;\rho_D]
  &= \mathrm{Tr}\Big[\rho_D(t_0){\cal T}_{\cal C}
     \exp\Big\{i
     \int_{{\cal C},x}\,\Phi_a(x)J_a(x)
     \Big\}\Big].
\end{align} 
Here, ${\cal T}_{\cal C}$ denotes time ordering along a closed time path or contour $\cal C$ which is discussed in more detail in \Sect{CTP} below.
$\rho_D(t_0)$ is the normalized density matrix describing the many-body system at the initial time $t_0$, which, in general, does not have the equilibrium form $\rho_D^{(\mathrm{eq})}\sim\exp\{-\beta H\}$. 
$\Phi$ is the operator of a $2\cN$-component scalar quantum field, with its space-time arguments $x=(t,\xv)=(x_0,\xv)$.
The index $a$ enumerates the two components $\Phi_a$, corresponding to the real and imaginary parts of a single complex Bose field $\Phi$, or, to their independent combinations $\Phi$ and $\Phi^\dagger$, and it enumerates $\cN$ different fields describing, e.g., different internal hyperfine atomic states.

Any correlation function of a quantum many-body system can be derived from this by functional differentiation with respect to the classical source field $J_a(x)$ and, for a closed system, by subsequently setting $J\equiv 0$.
Of particular interest are the connected correlation functions or cumulants which are analogously obtained from the Schwinger functional
\begin{align}
\label{eq:SchwingerW}
  W[J;\rho_D]
  &= -i\ln Z[J;\rho_D].
\end{align} 
For example, the field expectation value and the connected two-point function or propagator follow as
\begin{align}
  \phi_a(x)
  &=\langle\Phi_a(x)\rangle 
  = \mathrm{Tr}[\rho_D(t_0)\Phi_a(x)]
  \nonumber\\
  &=\left.\frac{\delta W[J;\rho_D]}
             {\delta J_a(x)}\right|_{J=0},
\label{eq:MFphi}
  \\
\label{eq:Gconn}
  G_{ab}(x,y)
  &=
  \langle{\cal T}_{\cal C}\Phi_a(x)\Phi_b(y)\rangle
  -\phi_a(x)\phi_b(y)
  \nonumber\\
  &= -i\left.\frac{\delta^2W[J;\rho_D]}
                {\delta J_a(x)\delta J_b(y)}\right|_{J=0}.
\end{align}
%

\subsubsection{Schwinger-Keldysh closed time path}
\label{sec:CTP}
In Eqs.~\eq{NEgenFuncZ} and \eq{Gconn}, ${\cal T}_{\cal C}$ denotes time ordering along a closed time path (CTP) $\cal C$ introduced in Refs.~\cite{Schwinger1961a,Mahanthappa1961a,Keldysh1964a}.
The time path ${\cal C}={\cal C}(\infty)$ extends from the initial time $t_0$ to infinite positive times and back to $t_0$.  
We use the short-hand notation
\begin{align}
  \label{eq:CTPtau}
  \int_{{\cal C}(t),x}
  =\Big[\int_{t_0,{\cal C}^+}^t\mathrm{d}x_0-\int_{t_0,{\cal C}^-}^t\mathrm{d}x_0\Big]
  \int \mathrm{d}^dx,
\end{align}
where ${\cal C}^\pm$ implies that the times over which it is integrated are evaluated on the outward (backward) branch of the CTP.
Note, that in Eqs.~\eq{MFphi}, \eq{Gconn}, the infinite real-time contour ${\cal C}$ can be replaced by a contour ${\cal C}(\tau)$ extending from $t_0$ to some finite time $\tau>t_0$ and back to $t_0$.
${\cal C}(\tau)$ must contain the times of interest, i.e., $x_0$ and $y_0$, but does not need to extend to any later time $\tau>x_0$ and $\tau>t_\mathrm{max}=\mathrm{max}\{x_0,y_0\}$, in Eqs.~\eq{MFphi} and \eq{Gconn}, respectively.
For any time later than the maximum time argument $t_\mathrm{max}$ of the correlation function considered, the external currents can be set to zero, $J_a(x)\equiv0$, $x_0>t_\mathrm{max}$, such that the respective `$+$' and `$-$' contributions on the outward and backward branches of $\cal C$, respectively, cancel in $Z$, leaving ${\cal C}(t_\mathrm{max})$.  

Note also that, if all fields are evaluated at times on the `$+$' branch, the `$-$' branch of ${\cal C}(\tau)$ ensures the normalization of the generating functional $Z[0;\rho_D]=1$, i.e., unity of the trace of the density matrix.

The generating functional $Z[J;\rho_D]$ has a functional integral
representation which can be found by inserting, to the left and right of $\rho_D$, a complete set of eigenstates of the Heisenberg field operators at the initial time, $\Phi(x_0=t_0,\xv)|\varphi^\pm\rangle=\varphi_0^\pm(\xv)|\varphi^\pm\rangle$, with $\varphi_0^\pm(\xv)\equiv\varphi(t_0,\xv)$:
\begin{align}
\label{eq:NEgenFuncZphipm}
  &Z_\tau[J;\rho_D]
  = \int [\mathrm{d}\varphi_0^+][\mathrm{d}\varphi_0^-]
     \langle\varphi_0^+|\rho_D(t_0)|\varphi_0^-\rangle
  \nonumber\\
  &\quad\times
     \int_{\varphi_0^+,{\cal C}(\tau)}^{\varphi_0^-}{\cal D}\varphi
     \exp\Big\{i\Big[S_\tau[\varphi]+
     \int_{{\cal C}(\tau),x}\,\varphi_a(x)J_a(x)\Big]\Big\}.
\end{align} 
We distinguish the generating functionals for different CTP endpoints by the subscript $\tau$.
For $\tau\to\infty$ we regain the original definition \eq{NEgenFuncZ}, $Z_\infty\equiv Z$.
The functional integral 
\begin{align}
  \int_{\varphi^+,{\cal C}(\tau)}^{\varphi^-}{\cal D}\varphi
  =\int\prod_{a,\xv}\Big[\prod_{t_0<x_0\le\tau}\mathrm{d}\varphi^+_a(x)
   \prod_{t_0<x_0\le\tau}\mathrm{d}\varphi^-_a(x) \Big]
\end{align}
sums over all field configurations along the CTP ${\cal C}(\tau)$.
$S$ denotes the classical action, defined as 
\begin{align}
\label{eq:ClassActL}
  S_\tau[\varphi]
  &= \int_{{\cal C}(\tau),x}{\cal L}(x),
\end{align} 
where $\cal L$ is the Lagrangian density
\newcommand{\fnCTPAction}{\value{footnote}}
\footnote{We call $S$, \Eq{ClassActL}, an action although it is, strictly speaking, a difference of two actions, each of which involves a time integration along one part of the CTP.}.
We will, for the first part of the discussion, not specify this action any further.
We postpone its definition to \Sect{FlowEqs} in which we will also return to the non-relativistic Bose gas introduced in \Eq{MBHamiltonian}.

\subsubsection{Initial state}
\label{sec:IniState}
A general initial density matrix can be parametrized as
\begin{align}
\label{eq:IniDMParam}
  \langle\varphi_0^+|\rho_D(t_0)|\varphi_0^-\rangle
  &= {\cal N}\exp\{if_{\cal C}[\varphi]\},
\end{align} 
with a normalization $\cal N$ and the functional $f$ expanded in powers of the fields:
\begin{align}
\label{eq:fcalC}
  f_{\cal C}[\varphi]
  &= \alpha^{(0)} + \sum_{n=1}^\infty\frac{1}{n!}\int_{x_1...x_n}
    \alpha^{(n)}_{a_1...a_n}(x_1,...,x_n)
  \nonumber\\
  &\quad\times
    \prod_{i=1}^n\varphi_{a_i}(x_i).
\end{align} 
Here, the coefficients $\alpha^{(n)}_{a_1...a_n}(x_1,...,x_n)$ vanish identically for all times different from $t_0$  (cf., e.g., Ref.~\cite{Berges:2004yj}). 

In many practical cases it is sufficient to specify, at time $t_0$, only the lowest-order correlation functions.
If an initial state is fully determined by the mean field $\phi_a(x)$ and the propagator $G_{ab}(x,y)$, then the initial density matrix, Eq.~(\ref{eq:IniDMParam}), can be written as a Gaussian in the field $\varphi$, i.e., all $\alpha^{(n)}$ with $n\ge3$ vanish identically.

We emphasize that non-Gaussian initial density matrices can be as easily implemented in the framework we are going to discuss in the following.

\section{Implementation of dynamical flow}
\label{sec:RGDynamics}
%
\subsection{Variable closed time path}
\label{sec:RegGenFunc}
On the basis of the closed-time-path formulation of the generating functional for correlation functions as discussed above we will now introduce our real-time flow equation approach to non-equilibrium dynamics.
The first step will be to define, for a particular finite closed time path (CTP) ${\cal C}(\tau)$, cf.~\Eq{CTPtau}, a regulator function having the form of an additional source in the generating functional $Z$ which reduces $Z$ to the finite-CTP generating functional $Z_\tau$.
By a Legendre transform of $Z_\tau$, a one-particle irreducible effective action  $\Gamma_\tau[\phi]$ is found which depends, as $Z_\tau$, parametrically on the end point $\tau$ of the time contour.
This is the starting point for deriving a partial differential equation which describes the flow of $\Gamma_\tau$ with $\tau$.
As can be seen already from our above discussion of the CTP, this flow equation is equivalent to a time-evolution equation for the effective action, and therefore for all correlation functions derived from it, with time arguments smaller than or equal to the CTP's end point.
\begin{figure}[tb]
\begin{center}
\includegraphics[width=0.47\textwidth]{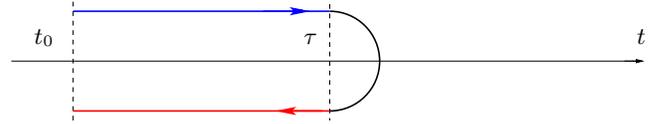}
\newline
\begin{picture}(0,0)(0,0)
\put(116,40){$t$}
\put(-112,40){$t_0$}
\put(-10,40){$\tau$}
\end{picture}
\end{center}
\vspace*{-3ex} \caption{(color online) The closed time path ${\cal
    C}(\tau)$ terminating at the time given by the parameter $\tau$.
  At later times, field fluctuations summed over in the generating
  functional do not contribute to Green functions the maximum time
  argument of which is $\tau$.}
\label{fig:CTPtau}
\end{figure}

Hence, the key idea of our approach to dynamics is to first consider the generating functional for Green functions where all times are smaller than a maximum time $\tau$. 
The time path ${\cal C}(\tau)$ closed at $t=\tau$, see \Fig{CTPtau}, leads to the generating functional in $Z_\tau$ \eq{NEgenFuncZphipm}. 
At $\tau =t_0$, this results in a trivial $Z_{t_0}$ where all information is stored in the initial density matrix $\rho_D(t_0)$.
From this initial condition, $Z_\tau$ can be computed by means of the
time evolution $\partial_\tau Z_\tau$ for all times $\tau>t_0$.  

\begin{figure}[tb]
\begin{center}
\vspace*{-7ex} 
\includegraphics[width=0.3\textwidth]{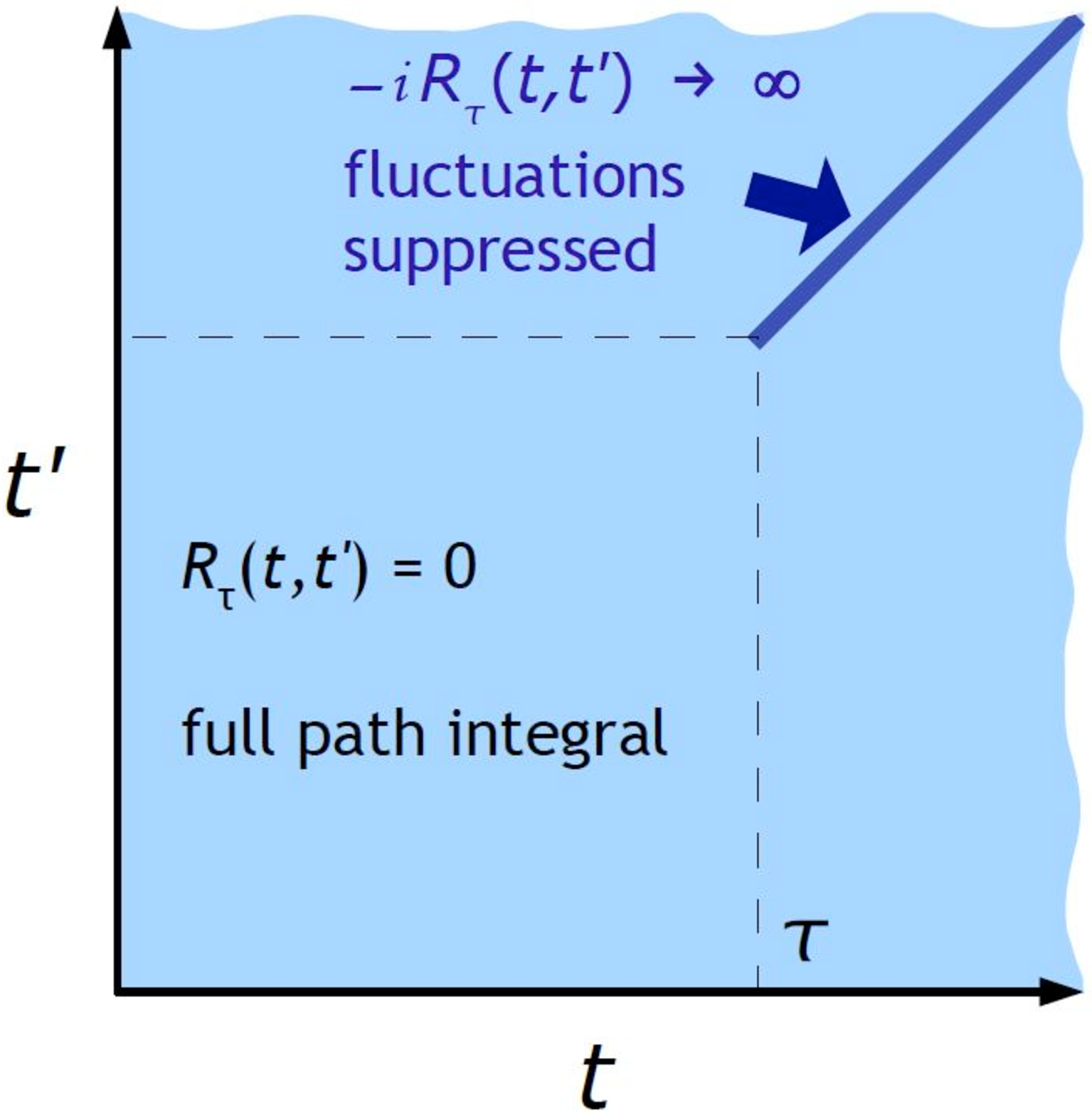}
\end{center}
\vspace*{-10ex} 
\caption{(Color online) The cut-off function $R_{\tau,ab}(x,y)$ in the time
plane $\{x_0,y_0\}=\{t,t'\}$, $t,t'\ge t_0$. The function vanishes identically 
except where $t=t'>\tau$, $\mathbf{x}=\mathbf{y}$, $a=b$ are fulfilled simultaneously where it tends to infinity and therefore 
implies a suppression of  
fluctuations in the generating functional at times greater than $\tau$.}
\label{fig:Rchoice}
\end{figure}

We will derive this evolution by using ideas from functional renormalization group (RG) theory \cite{Wetterich:1992yh,Bagnuls:2000ae,Litim:1998nf,Pawlowski:2005xe}.
To that end we note that $Z_\tau$ can be defined in terms of the full generating functional $Z_\infty\equiv Z$ in \eq{NEgenFuncZ} by suppressing the propagation for times greater than $\tau$. 
This suppression is achieved by introducing an additional two-point source $R_{\tau,ab}(x,y)$ which acts as a regulator cutting off the functional integration beyond $\tau$:
\begin{align}
\label{eq:defZtau}
  Z_\tau &= \exp\Big\{-\frac{i}{2} \int_{{\cal C},xy}\!
  \frac{\delta}{\delta J_a(x)}R_{\tau,ab}(x,y) \frac{\delta}{\delta
  J_b(y)}\Big\}Z\,.
\end{align}
The regulator function $R_\tau$ is chosen such that it suppresses the fields, i.e., $\delta/\delta J_a$, for all times $t> \tau$. 
This requirement does not fix $R_\tau$ in a unique way, and a simple choice 
is provided by 
\begin{equation}
\label{eq:Rchoice}
  -i R_{\tau,ab}(x,y)
  =\left\{\begin{array}{lcl} 
          \infty & \quad & \mathrm{for}\ x_0=y_0>\tau, 
	                   \mathbf{x}=\mathbf{y}, a=b \\[1ex] 
          0  	 & 		 & \mathrm{otherwise} 
          \end{array} \right.,
\end{equation}
see Fig.~\ref{fig:Rchoice}. 
We emphasize that the above choice fulfills the necessary requirements:
\begin{itemize}
\item $\lim_{\tau\to\infty} R_{\tau,ab}(x,y) = 0$ for all ($x_0,y_0$): 
This implies that the full quantum theory is recovered in the $\tau\to\infty$ limit, $\lim_{\tau\to\infty}Z_{\tau}[J;\rho_{D}] = Z[J;\rho_{D}]$.
\item $-iR_{\tau}$ is semi-positive definite, ensuring that $Z_{\tau}[J;\rho_{D}]$ does not diverge when derived from a normalized and finite $Z[J;\rho_{D}]$. 
Note that the following derivation of the flow equations can be achieved without functional integrals \cite{Pawlowski:2005xe} and that this condition is due to the approach presented here.
\item $-iR_{\tau,ab}(x,y)$ diverges at $\tau\to t_{0}$ for  $x_0=y_0$ greater than $t_0$, and $\mathbf{x}=\mathbf{y}$, $a=b$. 
This guarantees that for $\tau=t_0$ all fluctuations are suppressed and that $Z_{t_0}$ reduces to a trace over the initial density matrix $\rho_D(t_0)$, see \Eq{NEgenFuncZphipm}.
In this case all $n$-point functions derived from $Z_{t_0}$ are defined at $t_0$ only and reduce to the free classical ones, or any other physical boundary condition we can specify as discussed further below.
\end{itemize}
%

We emphasize that the cut-off $R_\tau$ in \eq{defZtau} suppresses any time evolution at times greater than $\tau$.
Correlation functions derived from $Z_\tau$ and therefore connected correlation functions obtained from
\begin{align}
\label{eq:SchwingerWtau}
  W_\tau[J;\rho_D]
  &= -i\ln Z_\tau[J;\rho_D]
\end{align} 
vanish as soon as at least one of their time arguments is larger than $\tau$.
For example, the connected two-point Green function obeys
\begin{align}
  \label{eq:Gproperties}
  G_{\tau, ab}(x,y) 
  &=\left\{\begin{array}{lcl} 
          0 & \quad 	& \mathrm{for}\ \mathrm{max}\{x_0,y_0\}>\tau \\[1ex] 
          G_{ab}(x,y)	& 		 & \mathrm{otherwise}
          \end{array} \right. .
\end{align}
Here and in the following, the index $\tau$ implies that a correlation function is obtained from a regularized generating functional, corresponding to ${\cal C}(\tau)$. 
In particular, $G_\tau$ is derived from $W_\tau$ in the same way as $G=G_\infty$ from $W$, cf. \Eq{Gconn}.

We point out that, in contrast to usual renormalization-group flows in the momentum domain, causality prevents the influence of times greater than $\tau$ on the dynamics up to time $\tau$. 
Hence, the sharp cut-off \eq{Rchoice} is the most physical one in the sense that it corresponds to integrating out all fluctuations being relevant for the evolution up to a particular time. 

Before we proceed to the derivation of flow equations we briefly discuss the possibility of alternative and additional regulators.
More general regulators $R_{\tau,ab}(x,y)$ in the time domain may be advantageous in other cases , e.g., when a cut-off is set in the relative-time, i.e., $(x_0-y_0)$-direction, see Fig.~\ref{fig:Rchoice}. 
Note that such a case would correspond to choosing a regulator in the frequency domain \cite{Canet:2003yu} which is conceptually closer to the traditional RG approach with cut-offs in the momentum domain.
The form of such an alternative temporal cut-off could be optimized along the lines described in Refs.~\cite{Litim:1998nf,Pawlowski:2005xe}.

We also remark that additional regulators in the spatial or momentum domains may be required to remove divergences through renormalization.
This can be done along the lines discussed in detail in Refs.~\cite{Wetterich:1992yh,Bagnuls:2000ae,Pawlowski:2005xe}.

\subsection{Effective action on finite closed time path}
\label{sec:RegEffAction}
Summarizing our above discussion, the time evolution of the correlation functions of interest can be derived from a flow equation for $Z_\tau$ as a function of $\tau$.
It will be more convenient, however, to work with the $\tau$-dependent effective action defined by the Legendre transform
\begin{align}
  \label{eq:effAction}
  \Gamma_\tau[\phi;\rho_D]
  &= W_\tau[J;\rho_D]-\int_{{\cal C},a}\phi_aJ_a
  - \frac{1}{2}\int_{{\cal C},ab}\phi_a  R_{\tau,ab}\phi_b.
\end{align}
Here, space-time arguments are suppressed, and we have extended the notation \eq{CTPtau} to include multiple integrals,
\begin{align}
\label{eq:defintCab}
  &\int_{{\cal C}(\tau),ab...}M_{ab...}N_{ab...}
  = \int_{{\cal C}(\tau),x_a}\int_{{\cal C}(\tau),x_b}\cdots 
  \nonumber\\
  &\quad\times\
     M_{ab...}(x_a,x_b,...)N_{ab...}(x_a,x_b,...)\,,
\end{align}
where $M$ and $N$ are arbitrary functions whose indices $a,b,...$ and space-time arguments $x_a,x_b,...$ are contracted. 
Possible further indices and arguments are not denoted explicitly.

\subsection{Correlation functions}
\label{sec:taudependentCorrFuncs}
In \Eq{effAction}, $\phi_a$ is the classical field expectation value,
\begin{align}
  \label{eq:DefphifromWtau}
  \phi_a(x)
  =W_{\tau,a}^{(1)}(x)[0;\rho_D]
  \equiv\delta W_\tau[J;\rho_D]/\delta J_a(x)|_{J\equiv0},
\end{align}
and this relation is to be inverted in order to determine $J=J[\phi;\rho_D]$ used in \Eq{effAction}.
Besides the field $\phi$ we will in particular be interested in the $\tau$ dependent connected Green function
\begin{align}
  \label{eq:DefGfromWtau}
  G_{\tau,ab}(x,y)
  &=-iW_{\tau,ab}^{(2)}(x,y)[0;\rho_D]
  \nonumber\\
  &=-i\left.\frac{\delta^2 W_\tau[J;\rho_D]}
                 {\delta J_a(x)\delta J_b(y)}\right|_{J=0}.
\end{align}
The regulator-dependent effective action functional \eq{effAction} is the generating functional of the regulator-dependent proper or one-particle irreducible (1PI) $n$-point functions
\begin{align}
\label{eq:DefGammataun}
  \Gamma^{(n)}_{\tau,a_1...a_n}(x_1,...,x_n)[\phi;\rho_D]
  =\frac{\delta^n\Gamma_\tau[\phi;\rho_D]}
   {\delta\phi_{a_1}(x_1)\cdots\delta\phi_{a_n}(x_n)}.
\end{align}

We finally show that $\Gamma^{(2)}_\tau+R_\tau$ is the inverse of the propagator $-iG_\tau$:
From Eqs.~\eq{effAction}, \eq{DefGammataun} we obtain the relations
\begin{align}
\label{eq:Gammatau1}
  &\Gamma^{(1)}_{\tau,a}(x)
  = -J_a(x) - \int_{{\cal C},y}R_{\tau,ab}(x,y)\phi_b(y),
  \\
\label{eq:Gammatau2}
  &\Gamma^{(2)}_{\tau,ab}(x,y) +R_{\tau,ab}(x,y)
  = -\frac{\delta J_a(x)}{\delta \phi_b(y)},
\end{align}
where the functional dependences on $\phi$ and $\rho_D$ have been suppressed.
Taking a further derivative of \Eq{DefphifromWtau} gives
\begin{align}
\label{eq:dphidJ}
  iG_{\tau,ab}(x,y)
  =\frac{\delta \phi_b(y)}{\delta J_a(x)}.
\end{align}
Here, $J=0$ is taken only after the derivative of $\phi$.
Multiplying Eqs.~\eq{Gammatau2} and \eq{dphidJ} yields
\begin{align}
\label{eq:GtauinverseofGammatauR}
  &-i\int_{{\cal C},z}\left[\Gamma^{(2)}_{\tau,ac}(x,z)+R_{\tau,ac}(x,z)\right]
  G_{\tau,cb}(z,y)
  \nonumber\\
  &\quad=\ \delta_{ab}\delta_{\cal C}(x-y),
\end{align}
where $\delta_{\cal C}(x-y)=\delta_{\cal C}(x_0-y_0)\delta(\mathbf{x}-\mathbf{y})$ is the 4-dimensional delta distribution, in the time variable on the CTP ${\cal C}$.

\subsection{Dynamical flow equation for the effective action}
\label{sec:FlowEqGamma}
The following final step is to derive explicitly the differential equations describing the flow of the functionals $Z_\tau$ and $\Gamma_\tau$ with the cut-off parameter $\tau$.
We derive the flow equation for the Schwinger functional $W_\tau$ by inserting \Eq{defZtau} into the definition \Eq{SchwingerWtau} and using Eqs.~\eq{DefphifromWtau} and \eq{DefGfromWtau},
\begin{align}
\label{eq:dtauWtau}
  \partial_\tau W_\tau[J;\rho_D]
  &= \frac{1}{2}\int_{{\cal C},ab}\left(G_{\tau,ab}+\phi_a\phi_b\right)
     \partial_\tau R_{\tau,ab}.
\end{align}
The flow equation for $\Gamma_\tau$ is obtained from \Eq{effAction} as
\begin{align}
\label{eq:dtauGammatau}
  &\frac{d}{d\tau}  \Gamma_\tau[\phi;\rho_D]
   = \frac{d}{d\tau} W_\tau[J;\rho_D]
  \nonumber\\
  &\quad
   -\ \int_{{\cal C},a}\phi_a\partial_\tau J_a
   -\frac{1}{2}\int_{{\cal C},ab}\phi_a  (\partial_\tau R_{\tau,ab})\phi_b
  \nonumber\\
  &\quad
  -\ \int_{{\cal C},b}iG_{\tau,ab}\Big[J_b 
   + \int_{{\cal C},c}R_{\tau,bc}\phi_c\Big]\partial_\tau J_a
  \nonumber\\
  &= \frac{1}{2}\int_{{\cal C},ab}G_{\tau,ab}\,\partial_\tau R_{\tau,ab}.
\end{align}
In the first equation above we used \Eq{dphidJ}.
Terms involving functional derivatives of $\phi$ with respect to $J$, multiplied by $\partial_\tau J$, cancel each other according to \Eq{Gammatau1}, i.e.,
\begin{align}
  &\int_{{\cal C},b}iG_{\tau,ab}\Big[\Gamma_{\tau,b}^{(1)}+J_b 
   + \int_{{\cal C},c}R_{\tau,bc}\phi_c\Big]\partial_\tau J_a=0,
\end{align}
such that the total derivative of $\Gamma_\tau$ with respect to $\tau$ reduces to a partial derivative.
Using this and Eqs.~\eq{DefphifromWtau} and \eq{dtauWtau} when evaluating the derivative of $W_\tau$ gives the second equation in \eq{dtauGammatau}.
Finally, using \Eq{GtauinverseofGammatauR}, the flow equation \eq{dtauGammatau} can be written as
\begin{align}
  \label{eq:flowGamma}
  \partial_\tau \Gamma_\tau
  &= \frac{i}{2} \int_{{\cal C},ab}\!
  \left[\0{1}{\Gamma^{(2)}_\tau+R_\tau}\right]_{ab}
  \partial_\tau R_{\tau,ab}\, .
\end{align}
\Eq{flowGamma} represents the central result of Ref.~\cite{Gasenzer:2008zz}.  
{This equation has the form of Wetterich's equation \cite{Wetterich:1992yh} for the exact functional renormalization group flow of the $k$-dependent effective average action $\Gamma_k$ with the momentum cut-off $k$.
Hence, it is technically similar to functional flow equations used extensively with regulators in momentum and/or frequency space to describe strongly correlated systems near equilibrium \cite{Wetterich:1992yh,Bagnuls:2000ae,Litim:1998nf,Pawlowski:2005xe}.
Its homogenous part relates to standard $\tau$-dependent renormalization \cite{Pawlowski:2005xe}, and has been studied, e.g., in Refs.~\cite{Boyanovsky:1998aa,Ei:1999pk}.
We point out that, in contrast to the usual RG flow equations  in momentum-frequency space, \Eq{flowGamma} describes a real-time flow and therefore the causal dynamical evolution of the quantum many-body system.
$\Gamma_{\tau}$ encodes the value of correlation functions with time arguments at and before $\tau$, for a given initial state characterized by $\Gamma_{t_{0}}$. 
}

\section{Dynamical flow equations for correlation functions}
\label{sec:FlowEqs}
%
\subsection{Model}
\label{sec:Model}
The discussion of our flow-equation approach to non-equilibrium dynamics has been, so far, very general, in that it was relying on only a few assumptions about the underlying (bosonic) quantum field theory and a specific choice for the regulator matrix $R$.
Extensions to and more specific formulations for, e.g., fermions and gauge bosons are straightforward and shall be postponed to future publications.
In the following we develop our approach further for the particular model of a bosonic many-body system with local quartic self-interactions.
We consider a quantum field theory for a complex $\cal N$-component field $\varphi_a(x)$  ($a=(i_{a},\alpha)$, $i_{a}=1,2$, $\alpha=1,...,{\cal N}$; $b=(i_{b},\beta)$, etc.) with quartic interactions  \footnotemark[\fnCTPAction],
\begin{align}
  S[\varphi]
  &= \frac{1}{2}\int_{{\cal{C}},x y} \varphi_a(x) iD_{ab}^{-1}(x,y)\varphi_b(y)
  \nonumber\\
  & -\frac{\lambda}{8{\cal N}} \int_{{\cal{C}},x} \varphi_a(x)\varphi_a(x)\varphi_b(x)\varphi_b(x) ,
\label{eq:Sclass}
\end{align}
This model describes, e.g., an ultracold Bose gas of atoms with $\mathcal{N}$ hyperfine sublevels whose interaction strength $g$ does not depend on the particular hyperfine scattering channel (`spins') of a pair of atoms.
In $d=3$ dimensions the coupling strength in such a system is $g={\lambda}/\mathcal{N}=4\pi a/m$,  $a$ being the $s$-wave scattering length.
Hence, the model possesses an $O(\mathcal{N})$ symmetry under transformations in spin space. 
The free inverse classical propagator of this model reads
\begin{align}
\label{eq:iinvG0NR}
  &iD^{-1}_{ab}(x,y)
  =\left.iG^{-1}_{0,ab}(x,y)\right|_{\phi=0}
  \nonumber\\
  &\quad=
   \delta_{\cal C}(x-y)\delta_{\alpha\beta}\left[-i\sigma^2_{i_{a}i_{b}}\partial_{x_0}
   -H_\mathrm{1B}(x)\delta_{ab}
     \right].
\end{align}
$\sigma^2$ is the Pauli 2-matrix.
We observe that \eq{iinvG0NR} is invariant under transformations $iD^{-1}\to UiD^{-1}U^\dagger$, with $U=\exp\{-i\sigma^2\alpha\}$, $\alpha\in\mathbb{R}$, such that $S_\tau[\varphi]$ is invariant under $U(1)$ rotations of the complex field $\varphi_{\alpha}=(\varphi_{1,\alpha}+i\varphi_{2,\alpha})/\sqrt{2}$.

$H_\mathrm{1B}(x)=-\nabla_{\mathbf{x}}^2/2m+V(x)$ is the one-body Hamiltonian of a particle of mass $m$ in a possibly time-dependent external (trapping) potential $V(x)$.

Our formulation carries straightforwardly over to, e.g., the relativistic Klein-Gordon theory with $\lambda\varphi^4$ interactions, in which the inverse propagator is fully diagonal and thus invariant under $O({\cal N})$ transformations ($a,b=1,\ldots,{\cal N}$):
\begin{align}
  iD^{-1}_{ab}(x,y)
  &=-\delta_{\cal C}(x-y)\delta_{ab}
    [\partial_{x_0}^2-\nabla_{\mathbf{x}}^2+m^2].
\label{eq:iinvG0Rel}
\end{align}
%

\subsection{Dynamical flow equations}
\label{sec:FlowGamman}
The flow equation \eq{flowGamma} for the full $\tau$-dependent effective action $\Gamma_\tau$ can be solved exactly only for special models.
This is not possible for the models introduced in the previous section if the spatial dimensionality is greater than zero.
The classical action for the above models is a polynomial in the fluctuating fields $\varphi_a$, with a single quartic interaction term.
One also expands the effective action $\Gamma_{\tau}$ in powers of the classical fields $\phi_a$ and restricts the discussion to the lowest moments $\Gamma^{(n)}_\tau$.
\begin{align}
  \Gamma_\tau[\phi] 
  &= \sum_{n=0}^{\infty} \frac{1}{n!} \int_{{\cal C},a_1\ldots a_n} 
    \ead{n}{\tau}{a_{1}\ldots a_{n}}(x_{1},\ldots, x_{n})[\bar\phi]
  \nonumber\\  
  &\quad\times\  
  \prod_{j=0}^{n} [\phi_{a_{j}}(x_j)-\bar\phi_{a_{j}}(x_j)]. 
\end{align}
Here, $\bar\phi_{a}$ is the expansion point, a possibly $\tau$-dependent classical field.
The natural choice is the solution $\bar\phi$ of the equations of motion, $\Gamma_\tau^{(1)}[\bar\phi]\equiv 0$, which extremizes the effective action $\Gamma_\tau$.
For this choice, the expansion starts effectively with the $n=2$ term as the linear term vanishes, and has maximal stability. 
For our model one solution exists where the field identically vanishes throughout space-time, $\bar\phi\equiv0$, and one or more where $\bar\phi$ is nonzero. 
In the following we will always assume that the field $\bar\phi$ is one of the solutions of the equation of motion.

Our goal is to describe the full time-evolution up to some time $t>t_0$, of the functions $\Gamma^{(n)}=\Gamma^{(n)}_{t}$, for $n\le4$.
In particular, we will be interested in the evolution of the connected two-point function $G=i[\Gamma^{(2)}]^{-1}=i[\Gamma^{(2)}_{t}]^{-1}$.
$G$ contains, e.g., the information on the evolution of the single-particle density matrix $n(\mathbf{x},\mathbf{y},t)=\langle\Phi^\dagger(\mathbf{x},t)\Phi(\mathbf{y},t)\rangle$ and the anomalous density matrix $m(\mathbf{x},\mathbf{y},t)=\langle\Phi(\mathbf{x},t)\Phi(\mathbf{y},t)\rangle$ (e.g. Refs.~\cite{Gasenzer:2005ze,Gasenzer2009a}). 

\subsubsection{Flow equation for $\Gamma^{(1)}_{\tau}$}
The proper one-point function satisfies the following flow equation, which is derived by functional differentiation of \Eq{flowGamma}, 
\begin{align}
  &\pdt \ead{1}{\tau}{a}[\phi] 
  = \frac{i}{2} \int_{{\cal C},cd} \ead{3}{\tau}{adc}[\phi] \,\grg{cd},
\label{eq:FlEqGamma1}
\end{align}
see also Fig.~\ref{fig:FlowEqs}.
The term in parentheses stands for the regularized line.
Here and in the following we use the abbreviations $\ead{n}{\tau}{a_1 \ldots a_n}= \ead{n}{\tau}{a_1 \ldots a_n}(x_{a_1}, \ldots, x_{a_n} )$, $\pr{\tau}{ab}=\pr{\tau}{ab}(x_a,x_b)$, $(MN)_{ab}=\int_{{\cal C},c}M_{ac}N_{cb}$, etc.
We point out that $\ead{1}{\tau}{c}$ appears in the equation of motion \eq{Gammatau1} which determines the dynamics of the field $\phi$.

\subsubsection{Flow equation for $\Gamma^{(2)}_{\tau}$}
The flow equation for the proper two-point function reads
\begin{align}
  \pdt \ead{2}{\tau}{ab}[\phi] 
  &= \frac{i}{2} \int_{{\cal C},cd} \ead{4}{\tau}{abcd}[\phi] \ \grg{dc} \nonumber \\
  &- \frac{1}{2} \int_{{\cal C},cdef} \Big[\ead{3}{\tau}{acd}[\phi]\,\pr{\tau}{de}\,\ead{3}{\tau}{bef}[\phi] 
  \nonumber\\
  &\qquad\times\ 
   \grg{fc} + P(a,b) \Big].
\label{eq:FlEqGamma2}
\end{align}

Here and in the following, $P(a,b,\ldots)$ stands for a sum over the respective previous terms, with all remaining permutations of the arguments $a,b,\ldots$. 
Note that the interchange of, e.g., $a$ and $b$ includes an interchange of the field indices and of the corresponding space-time arguments. 

To solve the above flow equations, initial correlation functions $\ead{n}{t_{0}}{a_{1}\ldots a_{n}}$ need to be specified which we discuss in \Sect{IniCondFlowEqs} below.

\subsubsection{Flow equations for coupling functions}
\begin{figure}[tb]
\begin{center}
\includegraphics[width=0.49\textwidth]{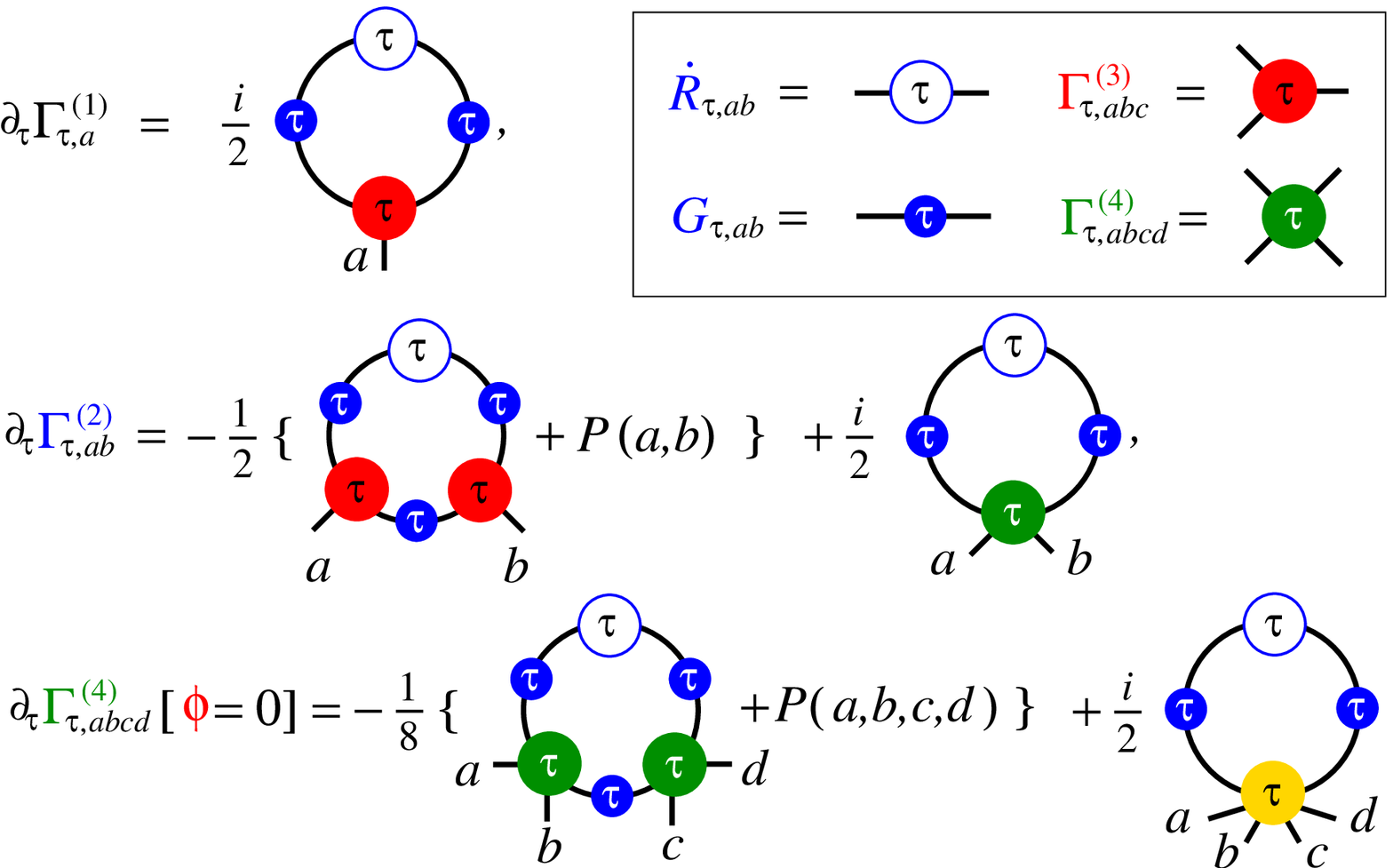}
\end{center}
\vspace*{-3ex} \caption{(color online) 
Diagrammatic representation of the general flow equations for $\Gamma_\tau^{(1)}[\phi],\Gamma_\tau^{(2)}[\phi]$, and $\Gamma_{\tau,abcd}^{(4)}[\phi=0]$, for a $\phi^4$-theory.  
Open circles with a $\tau$ denote $\partial_\tau R_{\tau,ab}$, solid lines with (blue) filled circles are $\tau$- and, in general, $\phi$-dependent two-point functions $G_{\tau,ab}=i[\Gamma_\tau^{(2)}+R_\tau]^{-1}_{ab}$.  
All other filled circles denote proper field-dependent $n$-vertices $\Gamma^{(n)}_{\tau,abcd}$, $n=3,4,6$. 
$P$ implies a sum corresponding to all permutations of its arguments.
The full flow equations for $\Gamma_{\tau,abc}^{(3)}[\phi]$ and $\Gamma_{\tau,abcd}^{(4)}[\phi]$ are given in App.~\ref{app:FlEqGamman}, \Fig{FlEqGamma34}.
  }
\label{fig:FlowEqs}
\end{figure}

The above flow equations need to be complemented with corresponding equations for the proper 3-point function $\ead{3}{\tau}{abc}[\phi]$ as well as the 4-point function $\ead{4}{\tau}{abcd}[\phi]$.
Fig.~\ref{fig:FlowEqs} shows a representation of the equations for the $\tau$-dependent proper functions $\Gamma_{\tau}^{(n)}$, for $n=1,2,4$ where the equation for $\ead{4}{\tau}{abcd}[\phi]$ is shown for $\phi\equiv0$.
In Appendix \ref{app:FlEqGamman} we provide the full flow equations for $\ead{3}{\tau}{abc}[\phi]$ and $\ead{4}{\tau}{abcd}[\phi]$ in diagrammatic form. 

Clearly, the equations for these functions involve proper functions of order $5$ and higher, and so on.
For practical computations, the resulting system of equations needs to be closed by truncation at some desired order, which of course is possible only at the expense of the exactness of the system.

We exemplify this here for the case of vanishing field expectation value.
For $\phi\equiv0$, the action (\ref{eq:Sclass}) implies that $\Gamma_{\tau}^{(3)}\equiv0$, and thus the flow of $\Gamma_\tau^{(1)}$ vanishes, see also Figs.~\fig{FlowEqs} and \fig{FlEqGamma34}.  
Hence, the equation for $\Gamma^{(2)}_{\tau}$ involves  only a term containing $\Gamma_\tau^{(4)}$,
\begin{align}
\label{eq:flowGamma2}
  \partial_\tau \Gamma^{(2)}_{\tau,ab}
  &= \frac{i}{2} \int_{{\cal C}}\!
  \Gamma^{(4)}_{\tau,abcd}
  (G_\tau\,[\partial_\tau R_\tau] G_\tau)_{dc}\, .
\end{align}
We need to supplement \Eq{flowGamma2} with the flow equation for $\Gamma^{(4)}_\tau$, 
which is illustrated in Fig.~\ref{fig:FlowEqs} (assuming (red) 3-point functions to vanish).
This system of equations is still exact. 
To close it we truncate by neglecting, in the equation for $\Gamma^{(4)}_\tau$, the term involving $\Gamma^{(6)}_\tau$,  
\begin{align}
\label{eq:flowGamma4}
\partial_\tau \Gamma^{(4)}_{\tau,abcd} 
  &= -\frac{1}{8}\int_{{\cal C}}\!  \Big\{
  \Gamma^{(4)}_{\tau,abef} G_{\tau,fg}
  \Gamma^{(4)}_{\tau,cdgh}
    \Big\}\,
  (G_\tau [\partial_\tau R_\tau] G_\tau)_{he} \,
  \nonumber\\
  &\quad
  +\, P(a,b,c,d).
\end{align}
Eqs. \eq{flowGamma2} and \eq{flowGamma4} represent a closed set of integro-differential equations for the proper functions up to fourth order. 
As we will show in the following, they allow to derive, for a particular cut-off time $\tau$, a set of dynamic equations describing the time evolution of the two- and four-point functions up to time $t=\tau$.  

\subsubsection{Initial conditions for flow equations}
\label{sec:IniCondFlowEqs}
For the solution of the coupled Eqs.~\eq{flowGamma2} and \eq{flowGamma4} initial functions $\ead{2}{t_{0}}{ab}$ and $\ead{4}{t_{0}}{abcd}$, respectively, need to be specified at $\tau=t_0$.  
We point out that, within the truncation scheme chosen above, we can insert any set of proper two- and four-point functions defined in their time arguments at and only at $t_0$, as long as we set all $n$-vertices for $n=1,3$, and $n>4$ to vanish.
For higher truncations more $n$-vertices can be specified at $t_{0}$ in an analogous manner. 
  
The truncation scheme discussed here corresponds to a Gaussian initial density matrix $\rho_D(t_0)$ since the four-point function has an influence on $\rho_D(t)$ only for $t>t_0$.
Here, we choose the respective classical proper functions defined by $S$ in
\Eq{Sclass}, 
\begin{align}
\label{eq:Gamma2t0}
  \Gamma^{(2)}_{t_0,ab}
  &=S^{(2)}_{ab}=iD^{-1}_{ab},
  \\
\label{eq:Gamma4t0}
  \Gamma^{(4)}_{t_0,abcd}
  &=S^{(4)}_{abcd}
  =-(\lambda/{\cal N})(\delta_{ab}\delta_{cd}+\delta_{ac}\delta_{bd}
     +\delta_{ad}\delta_{bc})
  \nonumber\\
  &\quad\times\ \delta_{\cal C}(x_a-x_b)\delta_{\cal
  C}(x_b-x_c)\delta_{\cal C}(x_c-x_d)\,,
\end{align}
and all higher $n$-vertices vanish.

\section{Dynamic Equations}
\label{sec:Dynamics}
\subsection{Causality}
\label{sec:Causality}
Causality implies that no physical observable measurable at time $t$ can depend on system properties arising at times larger than $t$.
This important physical fact distinguishes quantum field theory on the real time axis and the flow-equation approach from their formulation for equilibrium situations. 
Causality manifests itself, in the present theory, for the sharp cut-off regulator defined in \Eq{Rchoice} and the temporally local models introduced in \Sect{Model}, in the following properties of the correlation functions.
\begin{align}
\label{eq:Gtauab_theta}
  G_{\tau,ab} 
  &= i\eadi{\tau}{ab} \, \theta(\tau -t_a) \theta(\tau - t_b), 
  \\ 
\label{eq:GdtauG_theta}
  \grg{ab}  
  &=  \eadi{\tau}{ab} \, \partial_\tau \big[ \theta(\tau-t_a) \theta(\tau-t_b) \big].
\end{align}
For easier readability we shall use the notation $t_a=x_{a,0}$,  etc.
The $\theta$-functions are defined such that $\theta(\tau-t)$ is zero for $t > \tau$, otherwise it is 1. 
Relations similar to \Eq{Gtauab_theta} can be derived straightforwardly for all connected $n$-point Greens functions, see Appendix \ref{app:DerivationDynEqs} for the details.

We have chosen the sharp cut-off regulator \eq{Rchoice} in accordance with causility.
As discussed above, the flow equation for the generating functional $Z_\tau$, and therefore the flow equation for the effective action $\Gamma_\tau$ are time evolution equations for the respective functionals.
Hence, the dynamics of the proper vertices $\Gamma_\tau^{(n)}$ is readily given by the flow equations derived in the previous section.

At this point different avenues can be taken to solve these equations, see \Fig{Sketch}.
For the regulator \eq{Rchoice}, a convenient procedure is to integrate analytically the flow equations for the $\Gamma_\tau^{(n)}$, from initial time $t_0$ to some present time $t>t_0$.
The resulting equations have the form of integro-differential equations for the $n$-point functions, representing initial-value problems.
The form of these equations is familiar from other approaches discussed in the literature so far.
We adopt this approach in the present article.

\subsection{Dynamic equation for the field expectation value}
\label{sec:DynEqField}
{We begin by deriving the equation of motion for the field expectation value $\bar\phi_{a}(x_{a})$ defined 
by the solution of 
\begin{align}
\label{eq:Gamma1tc}
\ead{1}{\infty}{a}[\bar\phi]=  \ead{1}{t_a}{a}[\bar\phi]=0\,, 
\end{align}
Here we have used that the flow vanishes for $\tau \geq t_a=x_{a,0}$, see \eq{CausalityGammantequalsinfinity}. Note also, that then 
\eq{Gamma1tc} coincides with \eq{Gammatau1} at vanishing external current. In order to derive the dependence of the one-point function $\ead{1}{t_a}{a}[\phi]$ on its initial condition $\ead{1}{t_0}{a}[\phi]$ and the time evolution of the correlation, we decompose the effective action $\ea[t][\phi]$ as
\begin{equation} 
  \ea[t][\phi] 
  = \ea[t_0][\phi] 
  + \int\limits^t_{t_0} \mathrm{d}\tau\ \partial_{\tau}\ea[\tau][\phi],
\end{equation}
and calculate its functional derivative with respect $\phi_{c}$:
\begin{align} 
\label{eq:delGammtdelphic}
    &\frac{\delta\ea[t][\phi]}{\delta\cf{c}} 
    =  \frac{\delta}{\delta\cf{c}} \ \Big(\ea[t_0][\phi] 
      + \frac{i}{2}\int\limits^t_{t_0} \mathrm{d}\tau 
    \nonumber\\
    &\qquad\qquad\qquad\times
     \cii{ab} \left[\frac{1}{\eado{2}{\tau}[\phi]+R_{\tau}} \right]_{ab}
     \partial_{\tau} R_{\tau ,ab } \Big) 
     \nonumber \\
    &= \frac{\delta \ea[t_0][\phi]}{\delta\cf{c}} 
     + \frac{i}{2} \int\limits^t_{t_0} \mbox{d}\tau \cii{ab} \ 
       \ead{3}{\tau}{cab}[\phi] \grg{ba}.
\end{align}
Using the identities \eq{Gtauab_theta} and \eq{GdtauG_theta} we can carry out the integration over the cut-off time $\tau$. 
We obtain the dynamic equation for $\phi_{c}$ by inserting \Eq{delGammtdelphic} into \eq{Gamma1tc}:
\begin{equation}
\label{eq:EquationOfMotionField}
  \ead{1}{t_0}{c}[\bar\phi]
  +\frac{1}{2} \ci{t_c}{ab} \ \ead{3}{\tau_{ab}}{cab}[\bar\phi]  \ 
   \pr{\tau_{ab}}{ba}[\bar\phi]   
   =0.
\end{equation}
Here, we have introduced the notation
\begin{align}
\label{eq:tauab}
  \tau_{ab}&=\mathrm{max}\{t_a,t_b\}
\end{align}
as a shorthand for the maximum of time arguments of the two-point function.
We finally insert again the classical action \eq{Sclass} as initial condition to the integrated flow equation \eq{EquationOfMotionField} and arrive at the integro-differential equation of motion
\begin{align}
\label{eq:DynEqphi}
  &\Big(\ci{t_{c}}{d} iD^{-1}_{cd} \bar\phi_{d}\Big)
   - \frac{\lambda}{2\mathcal{N}}\bar\phi_d(x_{c})\bar\phi_d(x_{c})\bar\phi_c(x_{c})
   \nonumber\\
  &\quad=\ -\frac{1}{2} \ci{t_c}{ab} \ \ead{3}{\tau_{ab}}{cab}[\bar\phi]  \ 
   \pr{\tau_{ab}}{ba}[\bar\phi].
\end{align}
All time arguments of correlation functions involved in this equation are smaller than the present time $t_\mathrm{max}=t_{c}$.
This can be shown by considering the integral equation for $\ead{3}{\tau}{abc}$ provided in Appendix \ref{app:DerivationGamma34brokenphase}.
Hence, \Eq{DynEqphi} is causal and can be used with the set of equations derived in the following sections to solve the nonequilibrium time evolution of a system iteratively.

The time evolution of the field expectation value $\phi_c$ is also encoded in the dynamical flow of the one-point function \eq{FlEqGamma1}. We take the $\tau$-derivative of the equation of motion \eq{Gamma1tc} and arrive at 
\begin{equation}
  \label{eq:flowofminpre}
  \partial_\tau  \Gamma^{(1)}_{\tau,c}[\phi] 
  + \int_{\mathcal{C},d} \Gamma^{(2)}_{\tau,cd}[\phi] \partial_\tau \phi_d  =0\,.
\end{equation}
Solving \eq{flowofminpre} for $\partial_\tau \phi$ results in 
\begin{align}
\label{eq:flowofmin}
  \partial_\tau \phi_{c} 
  & = i  \ci{\tau}{d} G_{\tau,cd}\partial_\tau  \Gamma^{(1)}_{\tau,d}[\phi]
  \nonumber\\ 
  &=  -\ \frac{1}{2} \ci{\tau}{abd} G_{\tau,cd}  \ead{3}{\tau}{dab} \,\grg{ba}\,.
\end{align}
The $\tau$-evolution of $\phi$ is complementary to the time evolution of $\phi$ for times smaller than $\tau$. \Eq{flowofmin} 
completes the set of flow equations for $n$-vertices that is evaluated on the equation of motion. }

\subsection{Dynamic equations for propagator and $4$-vertex}
\label{sec:DynEqs}
%

\subsubsection{Dynamic equation for the propagator}
\label{sec:DynEqProp}
In this section we use the dynamical flow equation \eq{FlEqGamma2} to derive a integro-differential dynamic evolution equation for the proper 2-point function.
Integrating \Eq{FlEqGamma2} (recall that ${\cal C}={\cal C}(\infty)$) and employing Eqs.~\eq{Gtauab_theta} and \eq{GdtauG_theta} yields
\begin{align}
   \ead{2}{\tau}{ab}\Big|_{t_0}^{t} 
   &= \int\limits^{t}_{t_0} \mathrm{d}\tau \ \partial_{\tau} \ead{2}{\tau}{ab}  
     \nonumber \\
   &=  \frac{i}{2} \int\limits^{t}_{t_0} \mathrm{d}\tau 
      \ci{\infty}{cd} \ead{4}{\tau}{abcd} \ \grg{dc} 
    \nonumber \\
   &\qquad  - \frac{1}{2} \int\limits^{t}_{t_0} \mathrm{d}\tau \ 
     \Big[ \ci{\infty}{cdef} \ead{3}{\tau}{acd} \ \pr{\tau}{de} \ \ead{3}{\tau}{bef} \ 
     \nonumber\\
   &\qquad\qquad\times\grg{fc} + P(a,b) \Big] 
   \nonumber
\end{align}
\begin{align}
   \phantom{\ead{2}{\tau}{ab}\Big|_{t_0}^{t}}
   & = \frac{1}{2}  \ci{t}{cd} \ead{4}{\tau_{cd}}{abcd} \ \pr{\tau_{cd}}{dc} 
   \nonumber \\
   &\qquad + \frac{i}{2} \ci{t}{cdef} \ead{3}{\tau_{cdef}}{acd} \ \pr{\tau_{cdef}}{de} \ 
   \nonumber\\
   &\qquad\qquad\times\ead{3}{\tau_{cdef}}{bef} \ \pr{\tau_{cdef}}{fc}.
\label{eq:Gamma2t0t-derivation}
\end{align}
Here, we have used the notation \eq{tauab} and, analogously,
\begin{align}
\label{eq:tauabcd}
  \tau_{abcd}
  &=\mathrm{max}\{t_a,t_b,t_c,t_d\}
\end{align}
as a shorthand for the maxima of time arguments of the respective two- and four-point functions.
\eq{Gamma2t0t-derivation} is the dynamic equation for the two-point function $G_{ab}$ for the evolution from $t_{0}$ to $t$.
We will explain this in more detail in the following.

In the case of $t$ being smaller than the maximum \eq{tauab} of $t_a$ and $t_b$ the right hand side of \Eq{Gamma2t0t-derivation} must be zero because of the causality property \eq{CausalityFlowGamman}. 
Due to relation \eq{CausalityGammantequalsinfinity}, we can set the cut-off time to any value $t \geq t_a,t_b$.

From the above integrated flow equation we determine the integro-differential dynamic equation for the propagator by contracting it with $\texteadi{t}{ab}$:
\begin{align}
\label{eq:TwoPointFunctionDynamic}
  &\ci{\infty}{a'} \ead{2}{t_0}{aa'} \eadi{t}{a'b} 
  =  \cdf{a}{b}\, \delta_{ab}
  \nonumber\\ 
  &\qquad -\ \frac{1}{2} \ci{\infty}{a'} \ci{t}{cd} 
   \ead{4}{\tau_{cd}}{aa'cd}
   \pr{\tau_{cd}}{dc} 
   \eadi{t}{a'b} 
  \nonumber \\
  &\qquad -\ \frac{i}{2} \ci{\infty}{a'} \ci{t}{cdef} 
   \ead{3}{\tau_{cdef}}{acd} \pr{\tau_{cdef}}{de}
  \nonumber\\
  &\qquad\qquad\times\ 
   \ead{3}{\tau_{cdef}}{a'ef} \pr{\tau_{cdef}}{fc} 
   \eadi{t}{a'b}.
\end{align}
Our focus lies here on the microscopic theory defined by the classical action \Eq{Sclass} which is local in time.
Hence, the inverse classical propagator $\ead{2}{t_0}{ab}=iD^{-1}_{ab}$ is diagonal in its time arguments.
For a theory of this kind, the proper vertices satisfy \Eq{CausalityFlowGamman}, and thus the integration over $t_{a'}$ on the right hand side of \Eq{TwoPointFunctionDynamic} is restricted to the time path $\mathcal{C}(t)$. 
Choosing $t \geq t_b$ and using identity \eq{Gtauab_theta} we can replace $\eaditext{t}{a'b}$ by $-i\pr{t}{a'b}$ in the above equation.

Inserting, moreover, the initial condition \eq{Gamma2t0} we finally obtain the dynamic equation for the propagator in the form:
\begin{align}
\label{eq:PropagatorDynamic}
  &\ci{t}{a'} iD^{-1}_{aa'} \pr{t}{a'b} 
  =  i \cdf{a}{b}\delta_{ab}
  \nonumber\\
  &\qquad 
     -\ \frac{1}{2}  \ci{t}{a'cd} 
     \ead{4}{\tau_{cd}}{aa'cd} \pr{\tau_{cd}}{dc} \pr{t}{a'b} 
     \nonumber \\
  &\qquad - \frac{i}{2} 
     \ci{t}{a'cdef} 
     \ead{3}{\tau_{cdef}}{acd} \pr{\tau_{cdef}}{de} 
  \nonumber\\
  &\qquad\quad\times\
     \ead{3}{\tau_{cdef}}{a'ef}\pr{\tau_{cdef}}{fc} \pr{t}{a'b}.
\end{align}
The time derivatives in this integro-differential equation are embodied in $\ead{2}{t_0}{ab}=iG^{-1}_{0,ab}$, see Eqs.~\eq{iinvG0NR} and \eq{iinvG0Rel}.

\subsubsection{Integral equation for the vertex}
\label{sec:IntEqVertex}
The dynamic equation \eq{PropagatorDynamic} needs to be complemented with equations for the three- and four-point vertices.
In the following we will again consider first the special case that the field expectation value $\bar\phi$ vanishes, see \Sect{FlowGamman}, and present the result for the integrated equation for $\ead{4}{\tau}{abcd}$ for this simplified case.
In order to obtain a closed set of equations we furthermore truncate the equation for the vertex by dropping the diagram involving the six-point function, see \Fig{FlowEqs}.

The derivation proceeds in the same way as in \Eq{Gamma2t0t-derivation} for the proper two-point function and yields 
%
\begin{align}
\label{eq:Gamma4t}
  &\left.\Gamma^{(4)}_{\tau,abcd}\right|_{t_0}^t 
  =
  \frac{i}{2}\ci{t}{efgh}\!
  \Gamma^{(4)}_{\tau_{efgh},abef}G_{\tau_{fg},fg}
  \nonumber\\
  &\quad\times\
  \Gamma^{(4)}_{\tau_{efgh},cdgh}G_{\tau_{eh},he}
  +\, (a\leftrightarrow c) + (a\leftrightarrow d) .
\end{align}
The brackets denote terms with the given indices swapped.

\subsubsection{Implications from causality}
\label{sec:ImplicationsCausality}
Let us, at this point, come back to the consequences of the causality properties found in \Sect{Causality}, for the dynamic equation \eq{PropagatorDynamic} and \eq{Gamma4t}.
Without loss of generality we can assume $t_{a}\geq t_{b}$ since in the reverse case, the propagator can be obtained by symmetry $\prw{\tau}{ab}{x_a}{x_b} = \prw{\tau}{ba}{x_b}{x_a}$.

We thus can choose $t_{a}$ to coincide with the present time $t_\mathrm{max}$ at which \Eq{PropagatorDynamic} determines the further time evolution of the propagator, and set $t=\tau_{ab}=t_\mathrm{max}$.
Then, all time arguments appearing in the integrands on the right hand side of \Eq{PropagatorDynamic} are evaluated at times $\leq t_{\mathrm{max}}$.

The truncated dynamic equation \eq{PropagatorDynamic} then reads
\begin{align}
\label{eq:DynEqG2}
  &\ci{\tau_{ab}}{a'} iD^{-1}_{aa'} \pr{\tau_{ab}}{a'b} 
  =  i \cdf{a}{b}\delta_{ab}
  \nonumber\\
  &\qquad 
     -\ \frac{1}{2}  \ci{\tau_{ab}}{a'cd} 
     \ead{4}{\tau_{cd}}{aa'cd} \pr{\tau_{cd}}{dc} \pr{\tau_{ab}}{a'b}.
\end{align}
Note that neither in \Eq{Gamma4t} nor in \eq{DynEqG2}, the flow parameter is any longer a free parameter.
It is fixed by the time arguments of the correlation functions.
In particular, the flow parameter of the propagator is everywhere set to the maximum of its time arguments and could be dropped, such that $G_{\tau}\equiv G$ is recovered as the full Green function.

The time arguments of the proper four-point function appearing in \Eq{DynEqG2} and determined by \Eq{Gamma4t} are also at most as big as the maximum time $\tau_{ab}=t_{\mathrm{max}}$.
In the integral equation \eq{Gamma4t} for the vertex the cutoff parameter is evaluated at the maximum time $\tau_{efgh}$ occuring in the loop.
Moreover, taking into account the relations \eq{CausalityGammantequalst0} and \eq{CausalityGammantequalsinfinity} following from causality, the cutoff parameters $\tau_{efgh}$ on the r.h.s.~of \Eq{Gamma4t} can be replaced by either $t_{0}$ or the maximum time of the respective four-point function.

This leads to a fully causal structure of the full set of truncated dynamic equations exhibiting the full strength of our approach:
The dynamic equations for the evolution from $t_{0}$ to $t_\mathrm{max}$ can be used to solve the time evolution of the system iteratively.
At the maximum time $t$ no information is needed to propagate the evolution than what has been computed in the past.
In the following chapter we explicitly show that it is this causal structure of our equations which allows to rederive from them, in an $s$-channel truncation, the nonperturbative dynamic equations known from the 2PI effective action in next-to-leading order of a large $\mathcal{N}$ approximation.

\section{Nonperturbative $s$-channel truncation}
\label{sec:NonPertDyn}
We are now in the position to put the formalism to work. In this section we evaluate the system of equations for $\phi$, $G$, and $\Gamma^{(4)}$ within a non-perturbative $s$-channel approximation. This allows us to derive a closed system of dynamic equations which are then solved numerically with presently available hard- and software technology and which goes substantially beyond Hartree-Fock mean-field and Quantum-Boltzmann kinetic equations.
We show, in particular, that the form of the equations for the propagator corresponds to that of the Kadanoff-Baym dynamic equations and that the $s$-channel approximation of the self-energy functional is identical to that derived from the 2PI effective action in next-to-leading order (NLO) of a $1/{\cal N}$ expansion \cite{Berges:2001fi,Aarts:2002dj,Berges:2004yj,Mihaila:2003mh,Berges:2002wr,Gasenzer:2005ze,Aarts:2006cv,Berges2007a,Gasenzer2009a}.

As above we consider the complex scalar model defined by the classical action \eq{Sclass} and discussed in more detail in \Sect{Model}.
Most of the discussion presented here is independent of the particular choice, both of $\cal N$ and $D^{-1}_{ab}$. 
We will focus on the case of ${\cal N}$ complex fields describing an ultracold Bose gas of atoms in ${\cal N}$ internal states with identical $s$-wave contact interactions. 
Nonetheless, the derivations presented will apply equally well to the relativistic scalar $O({\cal N})$-symmetric theory the free inverse propagator of which is given in \Eq{iinvG0Rel}.
Hence, our results concerning the $1/{\cal N}$-expansion of the 2PI effective action are directly relevant for the dynamical theories presented in the work cited above.

As before we consider Gaussian initial conditions, i.e., only the proper functions $\Gamma^{(n)}$ up to order $n=4$ are assumed to be non-zero at the initial time $t_{0}$ of the flow, see the discussion at the end of \Sect{FlowGamman}.
The truncation scheme we use was described in detail in the previous section and takes into account the flow of the proper functions up to $n=4$.

Before proceeding we consider the flow of the three-point vertex, see \Fig{FlEqGamma34}, which is proportional to the three-point vertex itself provided one neglects, as in our truncation, the five-point function.
For our model defined by the classical action \eq{Sclass} the bare three-point function is given by
\begin{align}
  &S^{(3)}_{abc} [\phi] 
   = -\frac{\lambda}{\mathcal{N}} \cdf{a}{b} \cdf{b}{c}
   \nonumber\\ 
  &\qquad\times\  \big[ 
     \delta_{ab}\phi_{c}(x_c)   
  + \delta_{ac}\phi_{b}(x_b)
  + \delta_{bc}\phi_{a}(x_a)  \big]
  \label{eq:S3t0}
\end{align}
and is therefore proportional to $\phi$. 
Thus, if we start with a field expectation value $\bar\phi=0$ and with $\ead{3}{t_0}{abc}[\bar\phi]  = S^{(3)}_{abc} [\bar\phi]$, the three-point function is identical to zero for all cut-off times. 
Since the time derivative of $\bar\phi$ is, according to \Eq{DynEqphi}, proportional to the three-point function and $\bar\phi$, the field $\bar\phi$ remains zero during the entire time evolution.
This property of the flow equations is a consequence of the $Z_{2}$ symmetry 
of the microscopic theory.
The effective action remains in the symmetric phase, i.e., $\bar\phi=0$, during the flow if it is initially in the symmetric phase. 
Thus $\bar\phi$ and all odd proper $n$-point functions remain zero.
The following discussion will first focus on the symmetric phase and later be extended to the symmetry-broken phase $\bar\phi\neq 0$.

\subsection{Dynamics in the symmetric phase}
\label{Dynamics in the symmetric phase}
%
\subsubsection{Dynamic equations}
In the $s$-channel approximation introduced above we are left with the closed set of flow equations for the proper two- and four-point functions, Eqs.~\eq{Gamma2t0t-derivation} and \eq{Gamma4t}, respectively.

\begin{align}
\label{eq:TwoPoint Function vanishing field}
  \ead{2}{\tau}{ab}\Big|^{t}_{t_0} 
  &=  \frac{1}{2} \ci{t}{cd} \ead{4}{\tau_{cd}}{abcd} \pr{\tau_{cd}}{dc},
  \\
\label{eq:FourPoint Function vanishing field}
  \left.\Gamma^{(4)}_{\tau,abcd}\right|_{t_0}^t 
  &=
  \frac{i}{2}\ci{t}{efgh}\!
  \Gamma^{(4)}_{\tau_{efgh},abef}G_{\tau_{fg},fg}\Gamma^{(4)}_{\tau_{efgh},cdgh}
  \nonumber\\
  &\quad\times\
  G_{\tau_{eh},he}
  +\, (a\leftrightarrow c) + (a\leftrightarrow d) .
\end{align}
The resulting dynamic equation for the connected two-point function $G$ is given in \Eq{DynEqG2}.

\subsubsection{Mean-field approximation}
We start by considering the mean-field or Hartree-Fock (HF) approximation.
The mean-field limit implies that the evolution equations are local differential equations, i.e., the propagation kernel of any correlation function is local in the time and spatial variables.  
We obtain the mean-field approximation by neglecting the flow of the four-point function.
Hence, $\Gamma^{(4)}_{\tau}$ keeps, for all $\tau$, its initial form set in \Eq{Gamma4t0}.
For this truncation, the dynamic equation for the propagator reduces to the well-known time-dependent HF equations \cite{Hartree1928a,
Gasenzer2009a}
\begin{align}
\label{eq:Hartree Fock equation}
  &\ci{\tau_{ab}}{c}iD^{-1}_{ac}({x_a},{x_c}) \prw{\tau_{ab}}{cb}{x_c}{x_b} 
  =  i\delta_{ab} \cdf{a}{b} 
  \nonumber \\
  & +\ 
  \frac{\lambda}{2\mathcal{N}} \, 
  \Big[\prw{t_a}{dd}{x_a}{x_a} \  \delta_{ac} 
     + \prw{t_a}{ac}{x_a}{x_a}  
  \nonumber \\
  &\qquad   +\ \prw{t_a}{ca}{x_a}{x_a}  \Big] \ \prw{\tau_{ab}}{cb}{x_a}{x_b}.
\end{align}
These are differential equations in the first time argument of the two-time Green function $G$.
They can, however, be reduced to differential equations for the time-diagonal functions $G_{ab}(x,y)$ with $x_{0}=y_{0}$, see Refs.~\cite{Gasenzer:2005ze,Gasenzer2009a}, i.e., the off-diagonal elements with $x_{0}\not=y_{0}$ are irrelevant for the evolution on the diagonal.
Usually, the  HF equations are derived by the use of the equation of motion for the two-point Green function, under the assumption that the four-point function is a product of two two-point correlators \cite{KadanoffBaym1995a}. 
This Gaussian decorrelation assumption is at the heart of the above neglection of the flow of $\Gamma^{(4)}$.
We emphasize that due to the missing two-particle (four-point) correlations, the life time of small excitations of the system is infinite, and therefore the HF equations do not account for scattering processes which could lead to equilibration.

\subsubsection{$s$-Channel approximation}
\label{Vanishing mean field 2PI-Approximation}
It is necessary to consider the flow of the four-point function to derive dynamic equations that contain real collision effects of the particles.
In this section we shall consider an $s$-channel approximation to the truncated flow equation of the four-point vertex $\Gamma^{(4)}$.
We will show that this approximation is equivalent to the widely used NLO $1/\mathcal{N}$ approximation of the 2PI effective action \cite{Berges:2001fi,Aarts:2002dj,Berges:2004yj,Gasenzer2009a}.
We emphasize that, however, the derivation of these equations within our approach does not require a perturbative resummation of an infinite number of diagrams as it does within the 2PI approach.
Moreover, rederiving the dynamic equations in NLO of the 2PI $1/\mathcal{N}$ does not only reproduce an already known result but offers possibilities to obtain new insight into the nature of the approximation.

Within the present approach the $s$-channel approximation is based on the assumptions
\begin{align}
\label{eq:s-channel assumption}
  \ead{n>4}{\tau}{a_1 \ldots a_n} &\approx  0, 
  \\
\label{eq:s-channel assumption2}
  (1-P_{s}) \, \ead{m}{\tau}{a_1 \ldots a_m} &\ll P_{s} \,  \ead{m}{\tau}{a_1 \ldots a_m},  \end{align}
for $m=2$, $4$.  
Here, $P_{s}$ projects the proper two- and four-point functions onto the $s$-channel structure of the two- and four-point vertices, respectively. 
The flow-equation approach provides the possibility of a self-consistency check of the assumptions \eq{s-channel assumption} and \eq{s-channel assumption2}. 
This check involves calculating, using \Eq{FourPoint Function vanishing field}, the terms $(1-P_{s}) \, \ead{2}{\tau}{ab}$ and $(1-P_{s}) \, \ead{4}{\tau}{abcd}$, as well as $\ead{6}{\tau}{abcdef}$ that are generated by the $s$-channel proper two- and four-point functions. 
The above assumptions are self-consistent if the generated functions are sufficiently smaller than the corresponding $s$-channel terms. 

\begin{figure}[tb]
\vspace*{-3ex} 
\begin{center}
  \includegraphics[width=0.45\textwidth]{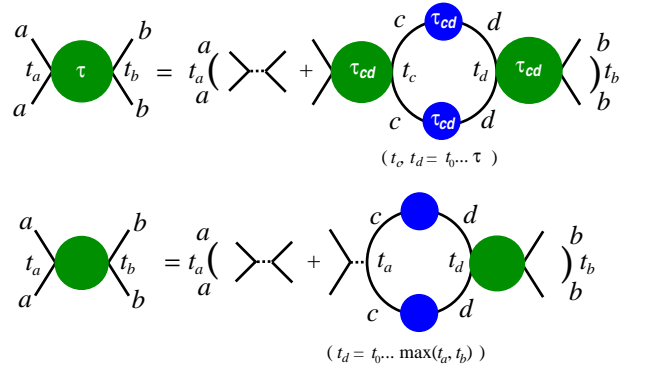}
\end{center}
\vspace*{-3ex} \caption{(color online) The upper equation is the
  $s$-channel projection of \Eq{Gamma4t} given in \Eq{FlowOf2PIVertex}.  The second
  equation defines the resummed vertex appearing in the
  Dyson-Schwinger equation derived from the NLO $1/{\cal N}$
  approximation of the 2PI effective action
  \protect\cite{Berges:2001fi}.  The two definitions are identical in
  every order of a perturbative expansion (see text).  Dashed lines
  denote the $s$-channel part of the bare vertex $\Gamma_{t_0,ab}^s$,
  see text.  All other symbols correspond to those in
  Fig.~\ref{fig:FlowEqs}.  Letters on internal lines indicate
  summation over field indices and integration over space and time
  (along the CTP from $t_0$ to the time given in parentheses and back).}
\label{fig:Bubblesum}
\end{figure}

We now define the $s$-channel approximation.
It takes into account, in addition to the Hartree-Fock terms, the flow of one scattering channel in the equation for the four-point function.
Labeling the field indices and space-time arguments of the four-point function $\Gamma^{(4)}$ in a unique way, the function has $s$-channel tensor structure if only elements with pairwise equal field indices and space-time arguments are nonzero. 
The possible combinations of pairs are usually referred to as the $s$-, $t$- and $u$-channels.
Any nonzero four-point function with three or more different indices or arguments is excluded from the $s$-, $t$-, and $u$-channel structures.
The four-point vertex in either of these channel projections is no longer fully symmetric under the interchange of any two of its indices and arguments.
Note that the classical four-vertex $\ead{4}{t_0}{abcd}=S^{(4)}_{abcd}$ which is given for our model in \Eq{Gamma4t0} consists of an $s$-, a $t$-, and a $u$-channel term, often called direct and exchange terms. 
While all three channels are taken into account on the mean-field level,  an $s$-channel truncation is chosen for the flow of the four-point function \Eq{FourPoint Function vanishing field}, by inserting
\begin{align}
\label{eq:Gamma4st0}
  \Gamma^{(4)s}_{t_0,abcd}
  &=-(\lambda/{\cal N})\delta_{ab}\delta_{cd}
  \nonumber\\
  &\quad\times\ \delta_\mathcal{C}(x_a-x_b)\delta_\mathcal{C}(x_b-x_c)
  \delta_\mathcal{C}(x_c-x_d),
\end{align}
into \Eq{FourPoint Function vanishing field}.
The integral equation for the $s$-channel vertex 
\begin{align}
\label{eq:Truncation4Vertex}
  &\tead{4}{\tau}{abcd}(x_a,x_b,x_c,x_d)
  = \delta_{ab}\delta_{cd}
  \nonumber\\
  &\quad\times\
   \delta_\mathcal{C}(x_a-x_b)\delta_\mathcal{C}(x_c-x_d) 
  \tv{\tau}{a}{c} 
\end{align}
reduces to
\begin{align}
\label{eq:FlowOf2PIVertex}
   \left.\tv{\tau}{a}{b}\right|_{t_0}^t
   &= \frac{i}{2} \ci{t}{eg} \tv{\tau_{eg}}{a}{e}\
   \pr{\tau_{eg}}{eg}(x_e,x_g) 
   \nonumber\\ 
   &\quad\times\ 
    \tv{\tau_{eg}}{g}{b} \ \pr{\tau_{eg}}{ge}(x_g,x_e). 
\end{align}
for a vertex function $\tv{t}{a}{b}$ which is a scalar in field-index space.
This integral equation is depicted in Fig.~\ref{fig:Bubblesum}, upper line. 

Before we discuss the structure of this equation further we provide the $s$-channel projection of the dynamic equation \eq{DynEqG2} for the two-point function $\pr{\tau}{ab}(x_a,x_b)$:
\begin{align}
\label{eq:s-channel final equation for propagator}
  &\ci{\tau_{ab}}{c}\left[
    iD^{-1}_{ac}({x_a},{x_c}) - \Sigma^{(0)}_{ac}(x_{a})\right]
    \prw{\tau_{ab}}{cb}{x_c}{x_b} 
  \nonumber \\
  &\quad= i\delta_{ab} \cdf{a}{b} 
  -\ i\frac{\lambda}{\mathcal{N}} \, 
  \ci{\tau_{ab}}{c}  \cfi{\tau_{ac}}{a}{c} 
  \nonumber \\
  &\qquad\times\ \prw{\tau_{ac}}{ac}{x_a}{x_c} \ \prw{\tau_{cb}}{cb}{x_c}{x_b} 
\end{align}
where
\begin{align}
\label{eq:Sigma0}
  &\Sigma^{(0)}_{ab}(x_{a}) 
  =  \frac{\lambda}{2\mathcal{N}} \, 
  \Big[\prw{t_a}{cc}{x_a}{x_a} \  \delta_{ab} 
     + \prw{t_a}{ab}{x_a}{x_a}  
  \nonumber \\
  &\qquad\qquad   +\ \prw{t_a}{ba}{x_a}{x_a}\Big]
\end{align}
is the HF mean-field contribution to the self-energy shift, cf. \Eq{Hartree Fock equation}, and 
\begin{equation}
\label{eq:nlo_schann_coupling_function}
 I_{t}(x_a,x_b) = -i \frac{\mathcal{N}}{\lambda} \  \tv{\tau}{a}{b}\Big|^{t}_{t_0}.
\end{equation}

Equation \eq{s-channel final equation for propagator} together with the definitions \eq{Sigma0} and \eq{nlo_schann_coupling_function}, and with the integral equation \eq{FlowOf2PIVertex} for the coupling function describe the time evolution of the propagator $G$ in the $s$-channel truncation of Eqs. \eq{TwoPoint Function vanishing field} and \eq{FourPoint Function vanishing field}.
These equations depend only on the initial conditions $D^{-1}$ and $\Gamma^{(4)}_{t_0}$ and can be expressed solely in $G_{ab}(x_{a},x_{b})=G_{\tau_{ab},ab}(x_{a},x_{b})$.
Nonetheless, the coupling function $\tv{\tau}{a}{b}$ depends, besides the arguments $x_{a}$, $x_{b}$ and the indices $a,b$, on the cut-off parameter $\tau$ which takes independent values, e.g., in \Eq{FlowOf2PIVertex}.

In Appendix \ref{app:sChanneltoNLO2PI1N} we show that the coupling function and therefore the above dynamic equations can be rewritten such that they get rid of this apparent additional degree of freedom and that they are equivalent to the dynamic equations derived from the 2PI effective action in NLO of a $1/{\cal N}$ expansion,
cf.~Ref.~\cite{Berges:2001fi} and \Fig{Bubblesum}, lower line.
The resulting coupling function is defined in Eqs.~\eq{Decomposition of the coupling function}, \eq{coupling function F in 2pi}, and \eq{coupling function rho in 2pi}.
The set of 2PI dynamic equations in NLO $1/\mathcal{N}$ approximation has been used extensively to investigate nonequilibrium dynamics and, expressed for $F$ and $\rho$, can be found, e.g., in Refs.~\cite{Berges:2001fi,Aarts:2002dj,Berges:2004yj,Gasenzer2009a}.

We emphasize that the proof builds crucially on the causal structure of the dynamic equations derived in our flow-equation approach.
This manifests the strength of the flow equation approach originating in causality as already pointed out in \Sect{ImplicationsCausality} above:
No information is needed for the propagation of the dynamic equations at time $t$ which has not been computed during the previous time evolution.
This renders the evaluation of the integral equations technically feasible.

Complementing the formal proof in Appendix \ref{app:sChanneltoNLO2PI1N} we show in the following that the above identity can be inferred in a comparatively easy way from the topology of the different terms in the flow equations for the two- and four-point functions: 
Consider the untruncated set of equations as displayed in \Fig{FlowEqs}.
First, non-$s$-channel contributions do not generate bubble-chains of the form shown in \Fig{Bubblesum}. 
Second, the term involving the six-point function to the flow of $\Gamma_\tau^{(4)}$ does not give rise to bubble-chains if inserted recursively into the first diagram on the right-hand side of the flow equation for $\Gamma_\tau^{(4)}$. To see this one has to generate the 
six-point function from the $s$-channel approximation for the four-point function. We emphasize that beyond the $s$-channel approximation the six-point term in the flow of the four-point function does create bubble diagrams. In turn, by dropping the second diagram with the six-point function and using the $s$-channel truncation, the iterated flow equation generates only bubble-chain diagrams with full propagators as lines.  
Hence, a $\tau$-integration of this set of flow equations leads to dynamic equations which include all bubble-chain contributions containing only bare vertices and full propagators and are therefore equal to those obtained to NLO in a $1/{\cal N}$ expansion of the 2PI effective action \cite{Berges:2001fi,Aarts:2002dj}.
We emphasize that the above topological arguments are generally valid when comparing resummation schemes inherent in truncated RG equations of the type of Eq.~\eq{flowGamma}, with those obtained from 2PI effective actions. 
This applies, e.g., to equilibrium flows \cite{Bagnuls:2000ae,Litim:1998nf,Pawlowski:2005xe} and thermal flows \cite{Litim:1998nf,Blaizot:2006rj}. 
For a comparison with 2PI results see Ref.~\cite{Blaizot:2006rj}, for the interrelation of 2PI methods
and RG flows Ref.~\cite{Pawlowski:2005xe}.
Note that decomposing the six- and higher $n$-point function according to their flow equation into diagrams containing only bare vertices and full propagators yields a loop expansion which, truncated at a particular order, is inequivalent to the expansion obtained without these higher-order vertices.
This is analogous to the general inequivalence of loop expansions derived from $n$PI effective actions with different $n$ \cite{Berges:2004pu}.

\subsection{Dynamics in the symmetry-broken phase, $\bar\phi\not=0$}
\label{Dynamics in the broken phase}
%
\subsubsection{Dynamic equations}
The complete flow equations for the one- and two-point functions are shown in \Fig{FlowEqs}.
In the case of spontaneously broken $U(1)$ symmetry, i.e., for a nonvanishing field expectation value $\bar\phi$, the three-point  vertex $\Gamma^{(3)}_{\tau}$ is in general non-zero.
Hence, we need to take into account the flow equations given in Appendix \ref{app:FlEqGamman}, see \Fig{FlEqGamma34}.
As before, we will consider only truncations, where we keep proper $n$-point function up to $n=4$, such that all diagrams will be omitted which contain a proper five- or six-point function.
This results in the dynamic equations for $\eado{3}{\tau}$ and $\eado{4}{\tau}$ given in App.~\ref{app:DerivationGamma34brokenphase}, see Eqs.~\eq{ThreePointFunction} and \eq{FourPointFunction}, respectively.
The dynamic equations for the propagator and the field expectation value were defined in Eqs.~\eq{PropagatorDynamic} and \eq{DynEqphi}, respectively.

\subsubsection{Mean-field approximation}
The dynamical Hartree-Fock-Bogoliubov (HFB) \cite{Hartree1928a,
Gasenzer2009a} mean-field equations are recovered by requiring the propagation kernels for $\phi_{a}$ and $G_{\tau_{ab},ab}$ to be local in the time and space variables.
For our model defined by the classical action \eq{Sclass} the bare three-point function $S^{(3)}_{abc}[\phi]$ is given by \Eq{S3t0} and is therefore proportional to $\phi$. 
Thus, the HFB dynamic equation for the field expectation value $\bar\phi$ is obtained by choosing
\begin{equation}
  \ead{3}{t}{abc}[\bar\phi]\equiv S^{(3)}_{abc}[\bar\phi]
\label{eq:Gamma3t0}
\end{equation}
and inserting this and $\ead{4}{t}{abcd}[\bar\phi]\equiv S^{(4)}_{abcd}$, see \Eq{Gamma4t0}, into  \Eq{DynEqphi}, giving
\begin{align}
\label{symmetrybrokenmeanfield}
   &\ci{t_{a}}{b} i D^{-1}_{ab}({x_a},{x_b}) \bar\phi_{b} - \frac{\lambda}{2\mathcal{N}} \bar\phi_{b}(x_a)\bar\phi_{b}(x_{a})\bar\phi_{a}(x_{a})
   \nonumber \\
  &=\ \frac{\lambda}{2\mathcal{N}} 
  \big[\prw{t_a}{bb}{x_a}{x_a}\bar\phi_{a}(x_{a})+\ \prw{t_a}{ab}{x_a}{x_a}\bar\phi_{b}(x_{a})
  \nonumber\\
  &\qquad+\ \prw{t_a}{ba}{x_a}{x_a}\bar\phi_{b}(x_{a})\big].
\end{align}
Setting the right-hand side of this equation to zero one recovers the classical equation of motion. 
For the free inverse propagator \eq{iinvG0NR} this is the Gross-Pitaevskii equation \cite{Gross1961a}; 
or the non-linear Klein-Gordon equation if $D^{-1}$ is given by \Eq{iinvG0Rel}.
The terms on the right hand side, involving the propagator $G$, are usually referred to as the back action of the excited modes onto the condensate within the HFB approximation.

The HFB evolution equation for $G$ is obtained by neglecting the term involving $\Gamma^{(3)}_{\tau}$ in \Eq{PropagatorDynamic} and inserting the bare four-point vertex \eq{Gamma4t0} for $\Gamma^{(4)}_{\tau}$, 
\begin{align}
\label{eq:s-channel final equation for propagator broken phase}
  &\ci{\tau_{ab}}{c}\left[
    i\priw{0}{ac}{x_a}{x_c} - \Sigma^{(0)}_{ac}(x_{a};\phi)\right]
    \prw{\tau_{cb}}{cb}{x_c}{x_b} 
  \nonumber \\
  &\qquad\qquad
  =\  i\delta_{ab} \cdf{a}{b} 
\end{align}
where
\begin{align}
\label{eq:Sigma0 broken phase}
  &\Sigma^{(0)}_{ab}(x_{a};\phi) 
  =  \Sigma^{(0)}_{ab}(x_{a}) 
  \nonumber\\
  &\quad+\ 
  \frac{\lambda}{2\mathcal{N}} \, 
  \Big[\bar\phi_{c}(x_a)^2 \  \delta_{ab} 
     + 2\bar\phi_{a}(x_a)\bar\phi_{b}(x_a)\Big].
\end{align}
As in the $U(1)$-symmetric phase with $\bar\phi=0$ the evolution of $G$ for $x_{a}\not=x_{b}$ is irrelevant for the evolution on the diagonal, $x_{a}=x_{b}$, such that the equations are entirely local and can be rewritten in the well-known form of the time-dependent HFB equations, see, e.g., Ref.~\cite{Gasenzer:2005ze,Gasenzer2009a}.

\subsubsection{$s$-Channel approximation}
\label{s-channel approximation vanishing field expectation value}
The dynamic equations in $s$-channel approximation are obtained, in analogy to the $U(1)$-symmetric case, by projecting, now, the proper $3$- and $4$-point functions onto the specific tensor structure, where both functions have at most two different external space-time variables $x_{a}$, $x_{b}$.
This projection was defined, for $\Gamma^{(4)}$, in \Eq{Truncation4Vertex}; for $\Gamma^{(3)}$ it reads 
\begin{align}
\label{three-point function in s-channel}
  &\tead{3}{\tau}{abc}(x_a,x_b,x_c) [\phi]
  \nonumber\\ 
  &\quad
  = \cdfw{a}{b} \cf{c} \tv{\tau}{a}{c}  
  \nonumber\\
  &\quad 
  +\ \cdfw{a}{c} \cf{b} \tv{\tau}{a}{b}
    \nonumber\\
  &\quad 
  +\ \cdfw{b}{c} \cf{a} \tv{\tau}{a}{b}.
\end{align}
The projection implies, in particular, that the four-vertices entering the full dynamic equation \eq{Gamma2t0t-derivation} of the propagator, i.e.,
$\ead{4}{\tau}{abcd}$ and $\ead{3}{\tau}{abe} \ \pr{\tau}{ef} \ \ead{3}{\tau}{fcd}$, both have, at most, two different external space-time variables. 
Hence, in the flow equations for the three- and four-point vertices shown in \Fig{FlEqGamma34}, besides the terms involving 5- and 6-point functions, all terms can be discarded with in general more than two external times.
In the equation for $\Gamma^{(4)}$, as before, only the second term on the right-hand side is kept, with two 4-point functions.

As far as the equation for $\Gamma^{(3)}$ is concerned, only the terms in the second line in \Fig{FlEqGamma34} are kept, with one three, and one four-point function, and thus, the fully evolved three-point function $\ead{3}{\tau_{abc}}{abc}$ is linear in $\phi_{d}(x)$ and a sum of terms with $(d,x)=(a,x_{a})$, $(b,x_{b})$, and $(c,x_{c})$.
As a result, the combination $\ead{3}{\tau}{abe} \ \pr{\tau}{ef} \ \ead{3}{\tau}{fcd}$ is quadratic in $\phi$, and in the different terms contributing to this vertex, the pair of fields is either connected by an internal line, or each is attached to one of the external lines, with different space-time variables.

Choosing the such defined truncation, one derives, as in the $U(1)$-symmetric case discussed in the previous section, coupled integro-differential equations for $\bar\phi$ and $G$ which are equivalent to the corresponding set of equations obtained from the 2PI effective action in NLO of a $1/\mathcal{N}$ approximation. 
These equations have been first derived, in Ref.~\cite{Aarts:2002dj} and can be found, e.g., for the case of the model defined in \Eq{Sclass}, in Ref.~\cite{Gasenzer2009a}.

\section{Conclusions}
\label{sec:Concl}

We have described in detail and extended the approach to far-from-equilibrium quantum field dynamics as put forward in Ref.~\cite{Gasenzer:2008zz}. 
The method is based on a time evolution of the system with respect to the flow of a maximum time marking the endpoint of the closed time path. 
In the present work we have shown that this approach directly implements the underlying causality of the time evolution and has practical applications beyond perturbation theory to describe far-from-equilibrium quantum dynamics. 
We have shown in detail that in an $s$-channel truncation the flow equations reduce to dynamic equations as known from a large-$\mathcal{N}$ expansion of the 2PI effective action. 
Although sub-leading terms in the expansion of the 2PI effective action are suppressed with additional powers of $1/\mathcal{N}$ they contain bare couplings, and, despite their resummation, the approximation seems formally questionable for strong couplings if $g/\mathcal{N}$ is large or at large times.
An important question is to what extent the NLO $1/\mathcal{N}$ approximation is applicable for large interaction strengths.
The present discussion so far is drawing mainly from benchmark tests for special-case systems \cite{Aarts:2001yn,Aarts:2006cv,Temme2006a}.  
Here we have shown that the self-consistency of such a truncation scheme can be evaluated in a closed form within the present approach by evaluating the flow of the six-point function as induced by the four-point function. 
We emphasize that such a closed self-consistency check goes beyond the evaluation of the NNLO contribution in a $1/\mathcal{N}$ expansion. 

Moreover, the equations provided in this work can be evaluated in an iterative manner, beyond the discussed $s$-channel truncation, without the need of resumming specific classes of diagrams.
This possibility resides on causality which implies that the propagation of the integro-differential equations in time requires only information computed during the previous evolution.
We hope to report on this matter in near future.

\acknowledgments \noindent 
The authors thank J. Berges, S. Borsanyi, S. Kehrein, D. Litim, H. Schoeller, D. Sexty, and C. Wetterich for useful discussions, and K.T. Mahanthappa for pointing out to them early work on the closed time path. 
They acknowledge support by the Deut\-sche Forschungsgemeinschaft, by the Helmholtz Alliance HA216/EMMI and the University of Heidelberg through the FRONTIER programme within the Excellence Initiative.
T.G. thanks KITP and the University of California at Santa Barbara, as well as M.~Holland, JILA, and the University of Colorado at Boulder for their hospitality, where part of this work was done. 
This research was supported in part by the National Science Foundation under Grant No. PHY05-51164.

\begin{appendix}

\section{Flow equations for proper $n$-point functions}
\label{app:FlEqGamman}
\begin{figure}[tb]
\begin{center}
\includegraphics[width=0.43\textwidth]{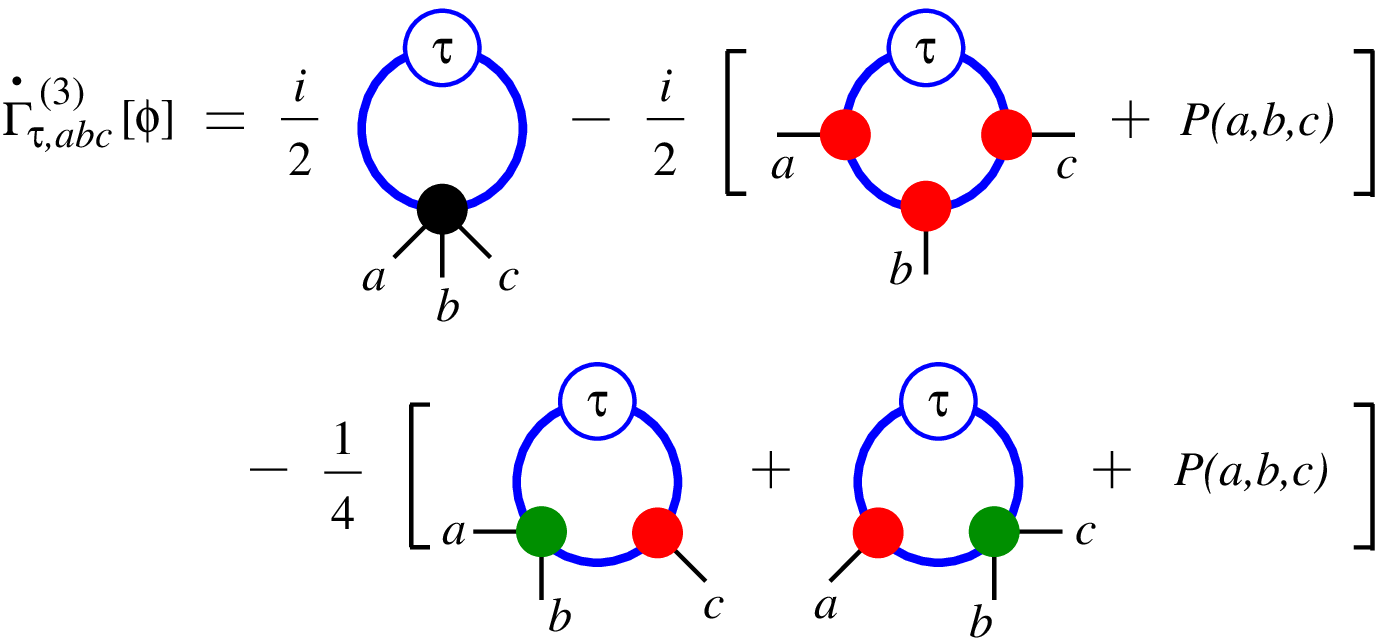}
\includegraphics[width=0.43\textwidth]{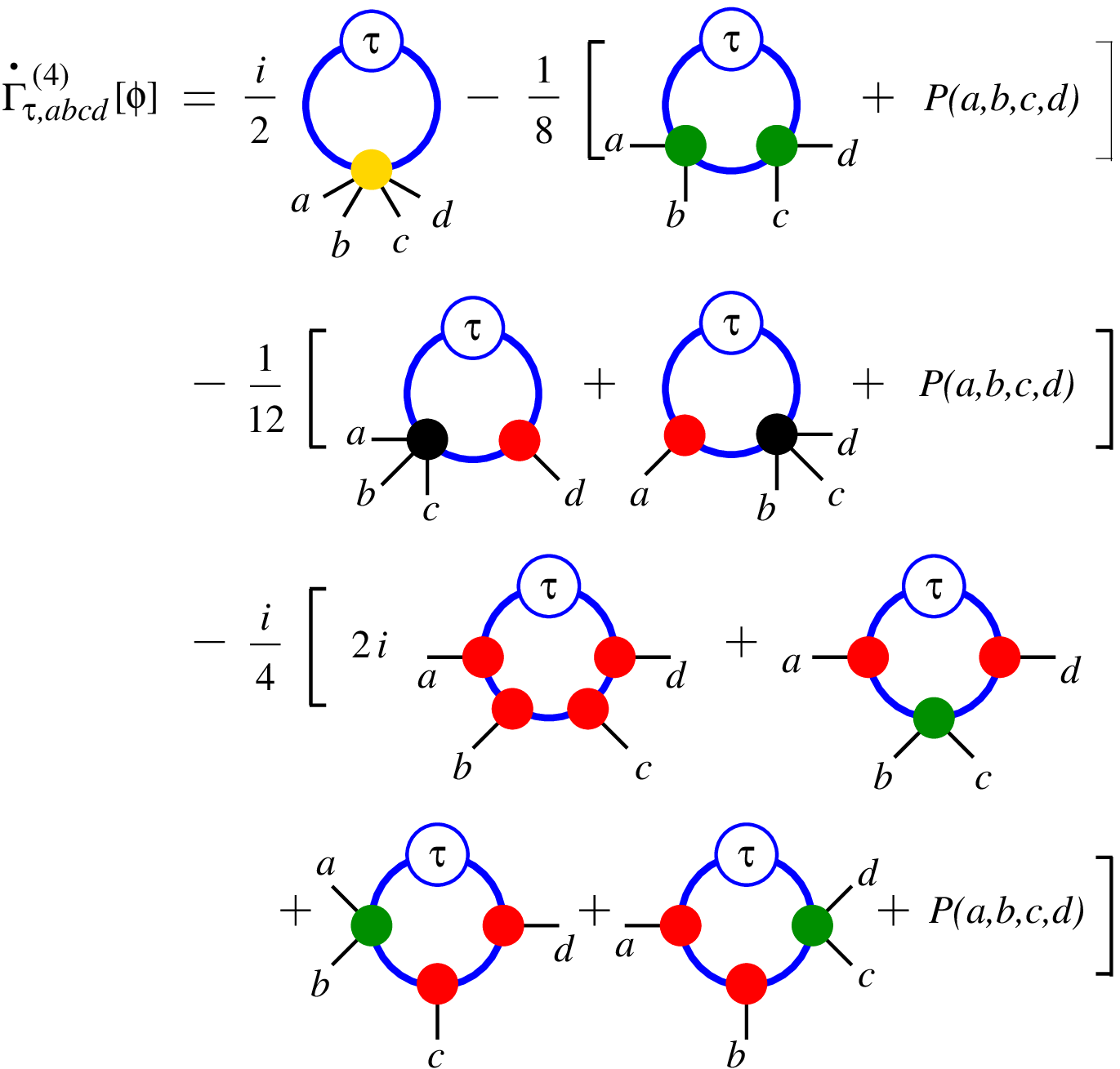}
\end{center}
\vspace*{-3ex} \caption{(color online) Diagrammatic representation of
  the general flow equations for
  $\Gamma_{\tau,abc}^{(3)}[\phi]$ and
  $\Gamma_{\tau,abcd}^{(4)}[\phi]$, for a $\lambda\phi^4$-theory.  
  Open circles with a $\tau$ denote $\partial_\tau R_{\tau,ab}$, solid (blue)
  lines are $\tau$- and, in general,
  $\phi$-dependent two-point functions
  $G_{\tau,ab}=i[\Gamma_\tau^{(2)}+R_\tau]^{-1}_{ab}$.  
  All other filled circles denote proper field-dependent $n$-vertices
  $\Gamma^{(n)}_{\tau,abcd}$, $n=3,4,6$, using the color scheme of \Fig{FlowEqs}. 
  $P$ implies a sum corresponding to all permutations of its arguments.
  }
\label{fig:FlEqGamma34}
\end{figure}

The flow equations for the one- and two-point proper functions as derived by functional differentiation of the exact flow equation \eq{flowGamma} with respect to the field $\phi$ are given in Eqs.~\eq{FlEqGamma1} and \eq{FlEqGamma2}, respectively.
In \Fig{FlEqGamma34} we provide the flow equations for $\eado{n}{\tau}[\phi]$, for $n=3,4$ and general $\phi$, in diagrammatic form, using the definitions of \Fig{FlowEqs}.

\section{Derivation of the dynamic equations}
\label{app:DerivationDynEqs}
In this appendix we derive the dynamic equations from the functional flow equations. 

\subsection{Implications from causality}
\label{app:PropertiesVertexFunctions}

The specific choice of the sharp cut-off regulator \eq{Rchoice} leads to an action functional $\ea[\tau]$ that includes all quantum fluctuations up to the cut-off time $\tau$ and reduces to the classical action for quantum fluctuations at times greater than $\tau$, see \Sect{RGDynamics}. For the proper $n$-point functions $\eado{n}{\tau}$ similar relations hold true: 

\textsl{Property 1:} 
For a microscopic model with bare vertices that are diagonal in time, combined with the sharp cut-off regulator \eq{Rchoice}, all proper $n$-point functions, with $n \geq 1$, obey 
\begin{equation}
  \pdt \ead{n}{\tau}{a_{1} \ldots a_{n}} (x_{1}, \ldots , x_{n}) = 0,
  \label{eq:CausalityFlowGamman}
\end{equation} 
for $\tau <t_{\max}$ with $t_{\rm max}=\max\{x_{1,0},...,x_{n,0}\}$. This encodes the causality of the flow. It follows immediately that 
\begin{equation}
  \ead{n}{\tau}{a_{1} \ldots a_{n}} (x_{1}, \ldots , x_{n}) 
  = \ead{n}{t_{0}}{a_{1} \ldots a_{n}} (x_{1}, \ldots , x_{n}). 
  \label{eq:CausalityGammantequalst0} 
\end{equation}
\Eq{CausalityFlowGamman} and \eq{CausalityGammantequalst0} follow directly from the flow equation \eq{flowGamma}. To that end we assume that \eq{CausalityGammantequalst0} is valid at a given flow time $\tau$, in particular it holds for $\tau=t_0$. 
The flow then maintains this property. The diagrammatic expansion of the flow equation, as e.g. shown in \Fig{FlEqGamma34}  for $n=3,4$, implies that at least one of the external times, $t_{\mathrm{max}}$, is greater than or equal to $\tau$ for any of the diagrams on the right hand side of the flow equation. 
With \eq{CausalityGammantequalst0} this particular leg renders the proper function it belongs to be equal to its bare counterpart at $\tau=t_{0}$. Since the bare vertices either vanish or are diagonal in the time variables, the diagram either vanishes, or at least one  time variable of the internal propagators attached to it is equal to $t_{\mathrm{max}}$.
Since all time integrations in the diagrams run from $t_{0}$ to, at most $\tau$, the respective propagators $G_{\tau,ab}(t_{\mathrm{max}},t_{b})$ vanish.
Since the other propagators in the diagram, the derivative of the regulator and the other vertices are bounded or have at most isolated singularities (This follows from the fact, that the bare vertices have only isolated singularities and that the flow in the cut-off time does not give rise to new singularities.) all diagrams on the right hand side of the flow equation vanish, which proves \Eq{CausalityFlowGamman}, and, by induction and in the limit $m\to\infty$, \Eq{CausalityGammantequalst0}.

\textsl{Property 2:} 
For the same kind of microscopic theory and regulator as above all proper $n$-point functions, with $n \geq 1$, obey \Eq{CausalityFlowGamman} if $\tau> t_{j}$ for all $j\in\{1,\ldots,n\}$, and 
\begin{align}
\label{eq:CausalityGammantequalsinfinity}
  \ead{n}{\tau}{a_{1} \ldots a_{n}} (x_{1}, \ldots , x_{n}) 
  &= \ead{n}{\infty}{a_{1} \ldots a_{n}} (x_{1}, \ldots , x_{n}),
\end{align}
if $\tau\ge t_{\rm max}$ with $t_{\rm max}=\max\{x_{1,0},...,x_{n,0}\}$. This relation states that the proper functions cease to flow once the cut-off parameter $\tau$ exceeds the maximum of the time arguments of the correlation function.
\Eq{CausalityGammantequalsinfinity} will not be necessary to derive the dynamic equations from the flow equations such that we can use the dynamic equations to derive \Eq{CausalityGammantequalsinfinity}.
We prove this for the example of the truncated equation  \eq{Gamma4t} for the four-vertex.
The time integrals on the right-hand side of the dynamic equation \eq{Gamma4t} run from $t_{0}$ to the cut-off time $t$ and back on the path $\mathcal{C}$.
The cut-off times $\tau_{efgh}$ appearing in the four-vertices on the right-hand side denote the maximum of the integration times and are thus smaller or equal to $t$.
Expanding \Eq{Gamma4t} by iteration, one arrives at a series of diagrams consisting of bare vertices $\Gamma_{t_{0}}^{(4)}$ and full propagators $G_{\tau_{ab}ab}$ with cut-off evaluated at the respective maximum of time arguments $\tau_{ab}=\mathrm{max}(t_{a},t_{b})$.
As will be shown in the following, in any of these diagrams the contributions from integrations of at least one time variable beyond the maximum $t_{\mathrm{max}}$ of external times vanish such that the cut-off parameter can be set to $t=t_{\mathrm{max}}$ which proves \Eq{CausalityGammantequalsinfinity}.
Consider a diagram with $\nu$ internal time integrations.
The integral over the $\nu$-dimensional volume of times in which at least one time $t_{j}>t_{\mathrm{max}}$ can be divided into integrals over subvolumes where the $\nu$ integration variables are ordered in a particular way $t_{P1}\le t_{P2}\le \ldots\le t_{P\nu}$ where $P$ denotes a permutation of the integers $1,...,\nu$.
Any of these subintegrals vanishes separately since the respective greatest time $t_{P\nu}$ is integrated, for any value of the second largest time $t_{P(\nu-1)}$, from $t_{P(\nu-1)}$ to $t$ and, along $\mathcal{C}$, back to $t_{P(\nu-1)}$.
Since the order of integrated as well as external times do not change along this integral, the time ordering in $G$ does not change and the outward and backward contributions along $\mathcal{C}$ cancel each other.

\subsection{$G_{\tau}$ and $G_{\tau}\dot{R}_{\tau}G_{\tau}$ for the sharp cut-off}
\label{app:GRdotG}
\label{app:Identities needed to derive the Dynamic Equations}
In this section we prove Eqs.~\eq{Gtauab_theta}, \eq{GdtauG_theta} which we use in Sect.~\ref{app:DerivationGamma4} to derive the dynamic equations of the proper $n$-point functions.
According to \Eq{GtauinverseofGammatauR} the propagator $G_{\tau}$ is the inverse of  $-i(\eado{2}{\tau} + R_{\tau})$.
We consider the case that $t_a,t_b < \tau$.
As before, $R_{\tau}$ is the sharp cut-off specified in \Eq{Rchoice}. 
Hence the integral $\ci{\infty}{c} R_{\tau,ac} \  \pr{\tau}{cb}$ vanishes since $R_{\tau, ac}$ is zero for all $t_c$ and the propagator is finite. 
Thus \Eq{GtauinverseofGammatauR} simplifies to
\begin{equation}
  \cdf{a}{b} \delta_{ab} 
  = -i \ci{\tau}{c} \ead{2}{\tau}{ac} \  \pr{\tau}{cb}
  \label{eq:GinvofGamma2belowtau}
\end{equation}
for $t_a,t_b < \tau$.
Here we also used that both the propagator and the proper two-point function are zero for $t_c > \tau$ because of Eqs.~\eq{Gproperties} and \eq{CausalityGammantequalst0}, respectively. 
Therefore the closed time path does not contribute for times $t_c$ greater than $\tau$, and from \Eq{GinvofGamma2belowtau} one has
\begin{equation}
\label{eq:form of propagator in appendix}
\pr{\tau}{ab}= i \eadi{\tau}{ab}  \ \mbox{for} \ \tau>\tau_{ab}. 
\end{equation}
Combining \Eq{form of propagator in appendix} with \Eq{Gproperties} we arrive at \Eq{Gtauab_theta}.

To derive the identity for $\grg{ab}$ we differentiate \Eq{GtauinverseofGammatauR} with respect to $\tau$, convolve with the propagator from the left.
Using again \Eq{GtauinverseofGammatauR} we obtain
\begin{align}
\label{eq:general expression for pdt g}
  i \pdt \pr{\tau}{ab} 
  =& -  \ci{\infty}{cd} \pr{\tau}{ac} ( \pdt \ead{2}{\tau}{cd} ) \pr{\tau}{db} 
  \nonumber\\
  &-\ \ci{\infty}{cd} \pr{\tau}{ac} ( \pdt R_{\tau,cd} )  \pr{\tau}{db}.
\end{align}
This relation holds for arbitrary regulator functions $R_{\tau}$.

Now, inserting \Eq{Gtauab_theta} into \Eq{general expression for pdt g} and substracting \Eq{Gtauab_theta} differentiated with respect to $\tau$, one arrives at \Eq{GdtauG_theta}, where one also uses the consequences of causality, \Eq{CausalityFlowGamman}, as derived in the previous section.

\subsection{The dynamic equation for $\Gamma^{(4)}_{\tau}$}
\label{app:DerivationGamma4}
\label{Derivation of the dynamic equation of the 4-point function}

In the following we present the derivation of the dynamic equation for the four-point vertex from the respective flow equation given in \Eq{flowGamma4}, for the sharp cut-off defined in \Eq{Rchoice}.
To this end we integrate \Eq{flowGamma4} over the cut-off time $\tau$ from $t_0$ to $t$ and obtain, after inserting relations \eq{Gtauab_theta} and \eq{GdtauG_theta} for $G$ and $G\dot R G$, respectively, 
\begin{align}
  &\ead{4}{\tau}{abcd}\Big|_{t_0}^{t} 
  = -\frac{i}{8} \ci{\infty}{efgh} \int_{t_0}^{t} \mbox{d}\tau \ \ead{4}{\tau}{abef} \left[ \eado{2}{\tau} \right]_{fg}^{-1} 
  \nonumber\\
  &\quad\times\ 
  \theta(\tau -t_{f}) \theta(\tau - t_{g}) \ead{4}{\tau}{cdgh} \left[ \eado{2}{\tau} \right]_{he}^{-1}
  \nonumber \\
  &\quad\times\
    \left[ \theta(\tau-t_{h})\delta(\tau-t_e) + \theta(\tau-t_{e}) \delta(\tau-t_h) \right] 
    \nonumber \\
  &\quad +\ P(a,b,c,d).
\end{align}
We use that $\ead{4}{\tau}{abcd}$ is symmetric under interchange of the $a,b,c,d$ and therefore the integral is invariant under the permutations $ (a,b,c,d) \rightarrow (b,a,c,d)$,  $ (a,b,c,d) \rightarrow (b,a,d,c)$, and  $ (a,b,c,d) \rightarrow (a,b,d,c)$, of space-time arguments as well as field indices.
Due to the cyclic property of the trace the permutation $ (a,b,c,d) \rightarrow (c,d,a,b)$ is identical to the renaming of the integration variables $(e,f,g,h) \rightarrow (g,h,e,f)$. The same is true for the permutations $(c,b,a,d) \rightarrow (a,d,c,b)$ and $(d,b,c,a) \rightarrow (c,a,b,d)$. 
Using this and performing the integration over $\tau$ we get
\begin{widetext}
\begin{align}
  \ead{4}{\tau}{abcd}\Big|_{t_0}^{t} 
  = -\frac{i}{2} \ci{\infty}{efgh} 
 \Big[ &\ead{4}{t_e}{abef} \left[ \eado{2}{t_e} \right]_{fg}^{-1} \ead{4}{t_e}{cdgh} \left[ \eado{2}{t_e} \right]_{he}^{-1} 
  \theta(t-t_{e})  \theta(t_e-t_f)  \theta(t_e-t_g)  \theta(t_e-t_h)  \nonumber \\
 +\ &\ead{4}{t_f}{abef} \left[ \eado{2}{t_f} \right]_{fg}^{-1} \ead{4}{t_f}{cdgh} \left[ \eado{2}{t_f} \right]_{he}^{-1}  
  \theta(t-t_{f})  \theta(t_f-t_e)  \theta(t_f-t_g)  \theta(t_f-t_h)  \nonumber \\
 +\ &\ead{4}{t_g}{abef} \left[ \eado{2}{t_g} \right]_{fg}^{-1} \ead{4}{t_g}{cdgh} \left[ \eado{2}{t_g} \right]_{he}^{-1}  
  \theta(t-t_{g})  \theta(t_g-t_e)  \theta(t_g-t_f)  \theta(t_g-t_h)  \nonumber \\
 +\ &\ead{4}{t_h}{abef} \left[ \eado{2}{t_h} \right]_{fg}^{-1} \ead{4}{t_h}{cdgh} \left[ \eado{2}{t_h} \right]_{he}^{-1}  
  \theta(t-t_{h})  \theta(t_h-t_e) \theta(t_h-t_f)  \theta(t_h-t_g) \Big] \nonumber \\
 +\ & (c,b,a,d) + (d,b,c,a),
\end{align}
\end{widetext}
where the last line denotes two different permutations of $(a,b,c,d)$ in the integral.
We define $\tau_{efgh}=\mathrm{max}(t_e,t_f,t_g,t_h)$ as the maximum of the integration times and simplify the above expression using that the sum of products of $\theta$-functions yields the integration volume $\ci{t}{efgh}$,
\begin{equation}
\label{rewriting the contour integral}
\ci{t}{a_{1} \ldots a_{n}} = \ci{\infty}{a_{1} \ldots a_{n}} \sum_{i=1}^{n} \tht{a_{i}} \prod_{j=1,j\neq i}^{n} \thf{a_{i}}{a_{j}}.
\end{equation}
Moreover, we use Eqs.~\eq{Gtauab_theta} and \eq{Gproperties} to write the integral in terms of the propagator $G$ and thus arrive at the form of the dynamic equation given in \Eq{FourPoint Function vanishing field}.
\\

\subsection{The dynamic equations for $\Gamma^{(3)}_{\tau}$ and $\Gamma^{(4)}_{\tau}$ for $\phi_{a}\not=0$}
\label{app:DerivationGamma34brokenphase}
The flow equations in the symmetry-broken phase, $\phi_{a}\not=0$, keeping at most the proper four-point function, are shown in Figs.~\fig{FlowEqs} and \fig{FlEqGamma34}, where the diagrams containing five- or six-point functions are omitted.
The derivation of the dynamic equations for the three- and four-point functions is performed as explained in the previous section. 
We obtain
\begin{widetext}
\begin{align}
\label{eq:ThreePointFunction}
  \ead{3}{\tau}{abc}\Big|^{t}_{t_0} 
  = & -\frac{1}{6}\ci{t}{defghi} \ead{3}{\tau'}{ade} \pr{\tau_{ef}}{ef} \ead{3}{\tau'}{bfg} \pr{\tau_{gh}}{gh} \ead{3}{\tau'}{chi} \pr{\tau_{id}}{id} 
  \nonumber \\
  & +\frac{i}{4} \ci{t}{efgh} \ead{4}{\tau^{*}}{abef} \pr{\tau_{fg}}{fg} \ead{3}{\tau^{*}}{cgh} \pr{\tau_{he}}{he} + P(a,b,c),
\end{align}
\begin{align}
\label{eq:FourPointFunction}
   \ead{4}{\tau}{abcd}\Big|^{t}_{t_0} 
   = & -\frac{i}{4} \ci{t}{efghijrs} \ead{3}{\tau''}{aij} \pr{\tau_{je}}{je} \ead{3}{\tau''}{bef} \pr{\tau_{fg}}{fg} 
                                                     \ead{3}{\tau''}{cgh} \pr{\tau_{hr}}{hr} \ead{3}{\tau''}{drs} \pr{\tau_{si}}{si}  
   \nonumber \\
   & -\ \frac{1}{4} \ci{t}{efghij} \ead{4}{\tau'}{abef} \pr{\tau_{fg}}{fg} \ead{3}{\tau'}{cgh} \pr{\tau_{hi}}{hi} \ead{3}{\tau'}{dij} \pr{\tau_{je}}{je} 
   \nonumber \\
   & +\ \frac{i}{16}\ci{t}{efgh} \ead{4}{\tau^{*}}{abef} \pr{\tau_{fg}}{fg} \ead{4}{\tau^{*}}{cdgh} \pr{\tau_{he}}{he} + P(a,b,c,d).
\end{align}
\end{widetext}
Here, we use the abbreviations $\tau^{*}=\tau_{efgh}$, $\tau'=\tau_{defghi}$, and $\tau'' = \tau_{efghijrs}$. 
The truncations made in equations \eq{ThreePointFunction} and \eq{FourPointFunction} are the omission of the five-point vertex in the dynamic equation for $\eado{3}{\tau}$ and the omission of the five- and six-point vertices in the dynamic equation for $\eado{4}{\tau}$.

\section{From the $s$-channel to the 2PI NLO $1/{\cal N}$ approximations}
\label{app:sChanneltoNLO2PI1N}

In this appendix we show that Eqs. \eq{FlowOf2PIVertex} to \eq{nlo_schann_coupling_function} obtained from an $s$-channel truncation of the dynamic equations derived in our flow-equation approach are equivalent to the dynamic equations derived from the 2PI effective action in NLO of a  $1/{\cal N}$ expansion. 
The proof does not rely on an expansion of the integral equations in a perturbative series of Feynman diagrams. 
The steps to prove the equivalence of the $s$-channel and the 2PI NLO $1/{\cal N}$ equations are the following:

\textsl{Step 1}:
We write the integral equation \eq{FlowOf2PIVertex} for the $s$-channel vertex in terms of the $s$-channel coupling function \eq{nlo_schann_coupling_function} and split the integral into three parts: 
The first part contains no coupling function $I_{\tau}$, the second part includes one coupling function and one bare vertex, and the third part embodies two $s$-channel coupling functions.

Making use of relations \eq{nlo_schann_coupling_function} and $\tv{t_0}{a}{b} = -({\lambda}/\mathcal{N}) \cdf{a}{b}$ we rewrite the integral equation \eq{FlowOf2PIVertex} as
\begin{align}
  &\cfi{t}{a}{b}
  = \frac{\lambda}{2\mathcal{N}} \ci{\infty}{cd} 
    \nonumber \\
  &\ \times\
  \{\tht{c} \thf{c}{d} + \tht{d} \thf{d}{c} \} 
  \nonumber \\
  &\ \times\
  [ i \cfi{\tau_{cd}}{a}{c} - \cdf{a}{c} ] \prw{\tau_{cd}}{cd}{x_{c}}{x_{d}} 
  \nonumber \\
  &\ \times\
  [ i \cfi{\tau_{cd}}{d}{b} - \cdf{d}{b} ] \prw{\tau_{cd}}{dc}{x_{d}}{x_{c}} .
\end{align}
We introduced the $\theta$-functions to extend the contour integral to infinity in order to reduce the number of different integrals. 
The above integral is now split into three terms: $I_{t}= I_{t}^{\delta\delta} + I_{t}^{\delta I} + I_{t}^{II}$, removing delta functions by integration, where %
\begin{align}
\label{eq:partIdd}
 I_{t}^{\delta\delta}
 &(x_{a},x_{b})
 =  \ \tht{a} \  \tht{b} \ \Pi_{\tau_{ab}}(x_a,x_b),
 \\
\label{eq:partIdI}
 I_{t}^{\delta I}
 &(x_{a},x_{b})
 = -i \ci{\infty}{c} \Big\{
 \cfi{\tau_{cb}}{a}{c} \Pi_{\tau_{cb}}(x_c,x_b) 
 \nonumber \\
 &\qquad\qquad 
 \times\ \Big[ \tht{c} \thf{c}{b} \thf{c}{a}
 \nonumber \\
 &\qquad\qquad\ \ 
 +\ \tht{b} \thf{b}{c} \thf{b}{a} \Big] 
 \nonumber \\ 
 &\qquad\qquad\quad 
 +\ \Pi_{\tau_{ac}}(x_a,x_c) \cfi{\tau_{ca}}{c}{b}
 \nonumber \\
 &\qquad\qquad 
 \times\ \Big[ \tht{a} \thf{a}{c} \thf{a}{b}
 \nonumber \\
 &\qquad\qquad\ \
 \ \ +\ \tht{c} \thf{c}{a} \thf{c}{b} \Big]  
 \Big\},    
\end{align}
\begin{align}
\label{eq:partIII}
 I_{t}^{II}
 &(x_{a},x_{b})
 = - \ci{\infty}{cd}  
 \cfi{\tau_{cd}}{a}{c} \, \Pi_{\tau_{cd}}(x_c,x_d) 
 \nonumber \\
 &\qquad\qquad\qquad\qquad 
 \times\  \cfi{\tau_{cd}}{d}{b} \,  
 \nonumber \\
 &\ \ 
 \times\ \Big[ \tht{d} \thf{d}{c} \thf{d}{b} \thf{d}{a} 
 \nonumber \\
 &\quad 
 \ +\ \tht{c} \thf{c}{d} \thf{c}{b} \thf{c}{a} \Big],
\end{align}
with the basic loop
\begin{align}
\label{eq:Gloop}
  \Pi_{\tau_{ab}}(x_a,x_b)
  &=\frac{\lambda}{2\mathcal{N}}\prw{\tau_{ab}}{cd}{x_a}{x_b}\prw{\tau_{ab}}{cd}{x_a}{x_b}.
\end{align}
To obtain the expressions \eq{partIdI} and \eq{partIII} we renamed integration variables to combine all terms into a single integral, and we used the property 
\begin{equation}
\label{eq:causalpropertyI}
  \cfi{t}{a}{b} = 0\quad 
  \mbox{for all}\quad t < \tau_{ab}
\end{equation}
of the $s$-channel coupling function to add theta functions as follows:
\begin{equation}
\label{eq:rewritten_Itau}
  \cfi{\tau_{cd}}{a}{c} = \left\{ \begin{array}{r@{\quad}l} \cfi{\tau_{cd}}{a}{c} \thf{c}{a}  
  & \mbox{for} \ t_c \geq t_d 
  \\[0.7ex] 
  \cfi{\tau_{cd}}{a}{c} \thf{d}{a} 
  & \mbox{for} \ t_d \geq t_c . 
  \end{array} \right.
\end{equation}
The property \eq{causalpropertyI} follows from the fact that all proper $n$-point vertices are equal to the respective bare vertex as long as one time argument is greater than the cut-off time, see \Eq{CausalityGammantequalst0}. 
\\

\textsl{Step 2}:
The propagator, i.e., the connected time-ordered Green function $\prw{\tau_{ab}}{ab}{x_a}{x_b}\equiv G_{ab}({x_a},{x_b})$ can be decomposed into the statistical and spectral correlation functions $\ff{\tau_{ab}}{ab}{a}{b}\equiv F_{ab}({x_a},{x_b})$ and $\rf{\tau_{ab}}{ab}{a}{b}\equiv \rho_{ab}({x_a},{x_b})$, respectively,
\begin{equation}
  \label{eq:decomposition of the propagator}
  G_{ab}({x_a},{x_b}) = F_{ab}({x_a},{x_b}) - \frac{i}{2} \sgn{a}{b} \rho_{ab}({x_a},{x_b}).
\end{equation}
This decomposition shifts the discontinuity due to the time-ordering of the noncommuting field operators into the signum function and thus yields two in general independent components, the statistical correlation function defined, for bosons, in terms of the anticommutator, $F_{ab}({x_a},{x_b})=\langle\{\Phi_{a}(x_{a}),\Phi_{b}(x_{b})\}\rangle_{c}/2$ and the spectral function defined in terms of the commutator of fields, $\rho_{ab}({x_a},{x_b})=i\langle[\Phi_{a}(x_{a}),\Phi_{b}(x_{b})]\rangle$.
The spectral function is related to the retarded propagator $G^R_{ab}(x_{a},x_{b})=\theta(t_{a}-t_{b})\rho_{ab}(x_{a},x_{b})$ as well as to the advanced Green function $G^A_{ab}(x_{a},x_{b})=-\theta(t_{b}-t_{a})\rho_{ab}(x_{a},x_{b})$.

We assume that we can analogously decompose the contour $s$-channel coupling function into two functions in real time by
\begin{equation}
\label{eq:Decomposition of the coupling function}
  \cfi{t}{a}{b}= \cfif{t}{a}{b} - \frac{i}{2} \cfir{t}{a}{b} \ \sgn{a}{b}.
\end{equation}
$\cfi{t}{a}{b}$ is symmetric under the interchange of $x_a$ and $x_b$ such that $\cfif{t}{a}{b}$ is symmetric and $\cfir{t}{a}{b}$ antisymmetric under $(x_a\leftrightarrow x_b)$. 
For the contribution $I^{\delta\delta}_{t}$ \Eq{Decomposition of the coupling function} is obvious from the decomposition of $\Pi_{t}$,
\begin{align}
  \label{eq:PiFt}
  \Pi^{F}_{\tau_{ab}}(x_{a},x_{b})
  &= \frac{\lambda}{2\mathcal{N}}\left(
  \spfr{\tau_{ab}}{a}{b}\right)
\end{align}
\begin{align}
  \label{eq:Pirhot}
  \Pi^{\rho}_{\tau_{ab}}(x_{a},x_{b})
  &=\frac{\lambda}{\mathcal{N}}
  \ff{\tau_{ab}}{cd}{a}{b} \rf{\tau_{ab}}{cd}{a}{b}.
\end{align}
For $I_{t}^{\delta I}$ we show this in Step 3 below, and $I_{t}^{II}$ vanishes identically as proven below.
 
We furthermore need that for any $\vec{x}_a$ and $\vec{x}_b$ one has 
\begin{align}
\label{eq:Irhovanishesateqtimes}
  \cfir{\tau}{a}{b}\big|_{t_a=t_b} 
  &= 0. 
\end{align}
For the spectral ($\rho$) part of $I_{t}^{\delta\delta}$ this follows from the relation $I^{\delta\delta,\rho}_{t}(x_a,x_b) = ({\lambda}/\mathcal{N}) \ff{\tau_{ab}}{cd}{a}{b} \rf{\tau_{ab}}{cd}{a}{b}$:
For the relativistic nonlinear Klein-Gordon model $\rf{t}{cd}{a}{b}|_{t_a=t_b} =0$.
For the Gross-Pitaevskii model $F$ is symmetric and $\rho$ antisymmetric under the exchange of field indices, such that the trace over the product $F\cdot\rho$ vanishes.
For $I^{\delta I}_{t}$ relation \eq{Irhovanishesateqtimes} is proven in Step 3.

In order to prove that $I^{II}_{t}$ vanishes identically we insert \Eq{Decomposition of the coupling function} into \Eq{partIII} and use  Eqs.~\eq{PiFt} and \eq{Pirhot} to obtain
\begin{widetext}
\begin{align}
&I^{II,F}_{t}(x_a,x_b) - \frac{i}{2}  I^{II,\rho}_{t}(x_a,x_b) \ \sgn{a}{b}
=  - \ci{\infty}{cd} \Big\{ \cfif{\tau_{cd}}{a}{c} - \frac{i}{2} \cfir{\tau_{cd}}{a}{c} \  \sgn{a}{c} \Big\} 
\nonumber \\
& \qquad\times\  
\Big[ \Pi^{F}_{\tau_{cd}}(x_{c},x_{d})   - \frac{i}{2} \Pi^{\rho}_{\tau_{cd}}(x_{c},x_{d})    \sgn{c}{d} \Big] 
\Big[ \cfif{\tau_{cd}}{d}{b} - \frac{i}{2} \cfir{\tau_{cd}}{d}{b} \ \sgn{d}{b} \Big] 
\nonumber \\
& \qquad\times\ 
\Big[ \tht{d} \thf{d}{c} \thf{d}{b} \thf{d}{a}  + \tht{c} \thf{c}{d} \thf{c}{b} \thf{c}{a} \Big].
\end{align}
\end{widetext}
In this expression only the signum functions depend on whether a particular integration time is on the forward or backward branch of the CTP. 
The theta-function terms allow to rewrite the time integrals along the CTP into integrals from $t_{0}$ to $t$. 
Eliminating the signum functions only terms with at least two $\rho$-components are left.
\Eq{Irhovanishesateqtimes}, the relation $\Pi^{\rho}_{\tau_{ab}}(x_{a},x_{b})  \big|_{t_a=t_b} = 0$, and the fact that $\thf{a}{b} \thf{b}{a}$ vanishes except for $t_{a}=t_{b}$ where it is finite then allow to show that the integrals over each of the remaining terms in the integrand vanish.

\textsl{Step 3}:
Next we consider the contribution $I^{\delta I}_{t}$ defined in \Eq{partIdI}.
Inserting \Eq{Decomposition of the coupling function} into \Eq{partIdI} we find its decompositions into contributions $I^{\delta I,F}_{t}$ and $I^{\delta I,\rho}_{t}$, confirming the decomposition \eq{Decomposition of the coupling function},
\begin{widetext}
\begin{align}
\label{eq:coupling function F in RG}
  \cfid{\delta I,F}{t}{a}{b} 
  &= \theta(t-\tau_{ab}) \Big[  
  - \int_{t_0}^{\mathrm{min}(t_a,t_b)} \mathrm{d}x_c 
  \Big\{ \thf{b}{a} \cfir{\tau_{bc}}{a}{c} \Pi^{F}_{\tau_{cb}}(x_{c},x_{b})
  - \thf{a}{b} \Pi^{F}_{\tau_{ac}}(x_{a},x_{c}) \cfir{\tau_{ca}}{c}{b} \Big\} 
  \nonumber \\
  &\qquad +\  \int_{t_0}^{\mathrm{max}(t_a,t_b)} \mathrm{d}x_c 
  \Big\{ \thf{b}{a} \cfif{\tau_{bc}}{a}{c} \Pi^{\rho}_{\tau_{cb}}(x_{c},x_{b}) 
  - \thf{a}{b} \Pi^{\rho}_{\tau_{ac}}(x_{a},x_{c}) \cfif{\tau_{ca}}{c}{b} \Big\} 
  \Big],
\end{align}
\begin{align}
  \label{eq:coupling function rho in RG}
  \cfid{\delta I,\rho}{t}{a}{b} 
  &= \theta(t-\tau_{ab}) \Big[    
  \thf{b}{a} \int_{t_a}^{t_b} \mathrm{d}x_c 
    \cfir{\tau_{cb}}{a}{c} \Pi^{\rho}_{\tau_{cb}}(x_{c},x_{b}) 
  -  \thf{a}{b} \int_{t_b}^{t_a} \mbox{d}x_c 
    \Pi^{\rho}_{\tau_{ac}}(x_{a},x_{c})  \cfir{\tau_{ac}}{c}{b} 
    \Big].
\end{align}
\end{widetext}
These equations are explicitly symmetric under the exchange of $x_{a}$ and $x_{b}$.
Moreover, we see that \Eq{Irhovanishesateqtimes} is valid for $I^{\delta I,\rho}_{t}$ as the integral limits are equal in this case.

In Eqs.~\eq{partIdd}, \eq{coupling function F in RG}, and \eq{coupling function rho in RG} the theta functions involving $t$ imply that the cut-off parameter $t$ of $I_{t}(x_{a},x_{b})$ can be set to $\tau_{ab}$.
Moreover, due to the theta functions in the integrands in Eqs.~\eq{coupling function F in RG} and \eq{coupling function rho in RG}, one can also set the cut-off parameters of the $I$ functions appearing in the integrands to the maximum of their respective time arguments.
Since the cut-off parameters $\tau$ of any of the functions $I_{\tau}$, $F_{\tau}$, and $\rho_{\tau}$ in the above definitions of $I^{F,\rho}_{\tau}$ are now fixed to the maximum of their respective time arguments, they can be neglected.
This allows to write the integral equations for $I^{F,\rho}_{\tau_{ab}}(x_{a},x_{b})\equiv I^{F,\rho}(x_{a},x_{b})$ as
\begin{align}
\label{eq:coupling function F in 2pi}
  I^{F}(x_a,x_b) 
  =&\ \Pi^{F}(x_a,x_b) 
  -\ \int_{t_0}^{t_a} \mathrm{d}x_c 
  I^{\rho}(x_a,x_c) \, \Pi^{F}(x_c,x_b)
  \nonumber \\
  & -\  \int_{t_0}^{t_b} \mathrm{d}x_c 
  I^{F}(x_a,x_c) \, \Pi^{\rho}(x_c,x_b),
\end{align}
and 
\begin{align}
  \label{eq:coupling function rho in 2pi}
  I_{\rho}(x_a,x_b) 
  =&\ \Pi^{\rho}(x_a,x_b) 
  -\ \int_{t_b}^{t_a} \mathrm{d}x_c 
  I^{\rho}(x_a,x_c) \Pi^{\rho}(x_c,x_b). 
\end{align}
To arrive at this form one uses that the integrals are invariant under the exchange $I^{\rho}(x_a,x_c) \Pi^{F}(x_c,x_b)=\Pi^{F}(x_a,x_c)I^{\rho}(x_c,x_b)$ etc.
The integral equations \eq{coupling function F in 2pi}  and \eq{coupling function rho in 2pi} are identical to the equations determining the coupling functions $I^{F,\rho}$ entering the self-energy derived from the 2PI effective action in next-to-leading order of a $1/\mathcal{N}$ expansion, see, e.g., Refs.~\cite{Berges:2004yj,Gasenzer2009a}.
\vfill

\end{appendix}


\end{document}